%% file: acmtecsgugg2014.tex
\documentclass[prodmode, acmtecs]{acmsmall}
\pdfoutput=1
\usepackage{graphicx}
\usepackage{subfigure}
\usepackage{amsmath}
\usepackage{amssymb}
\usepackage{amsbsy}
\usepackage{float}
\usepackage{graphics}
\usepackage{caption}
\usepackage{cite}
\usepackage{algorithmic}
\usepackage{courier}
\usepackage{xcolor}
\usepackage{listings}
\usepackage{url}
\usepackage{booktabs}
\input{myDefinitions.tex}
\newcommand{\revtwo}[1]{\textcolor[rgb]{0.00,0.00,0.00}{#1}}

\newcommand{\revfour}[1]{\textcolor[rgb]{0.00,.00,0.00}{#1}}
\newcommand{\revfive}[1]{\textcolor[rgb]{0.00,.00,0.00}{#1}}
\begin{document}

\markboth{Gugg et al.}{Real-Time Solution of Inverse Problems on Embedded Systems}

\title{An Algebraic Framework for the Real-Time Solution \\ of Inverse Problems on Embedded Systems}
\author{Christoph Gugg, Matthew Harker, Paul O'Leary and Gerhard Rath}

\begin{abstract}
This article presents a new approach to the real-time solution of inverse problems on embedded systems. 
The class of problems addressed corresponds to ordinary differential equations (ODEs) with generalized linear constraints, whereby the data from an array of sensors forms the forcing function.
The algebraic discretization of the problem enables a one-to-one mapping of the ODE to its discrete equivalent linear differential operator, together with an additional matrix equation representing the constraints.
The solution of the equation is formulated as a least squares (LS) problem with linear constraints.
The LS approach makes the method suitable for the explicit solution of inverse problems where the forcing function is perturbed by noise. 
The algebraic computation is partitioned into a initial preparatory step, which precomputes the matrices required for the run-time computation; and the cyclic run-time computation, which is repeated with each acquisition of sensor data. 
The cyclic computation consists of a single matrix-vector multiplication, in this manner computation complexity is known a-priori, fulfilling the definition of a real-time computation.
Numerical testing of the new method is presented on perturbed as well as unperturbed problems; the results are compared with known analytic solutions and solutions acquired from state-of-the-art implicit solvers. 
In all performed numerical tests the new method was both faster and more accurate for repeated solutions of the same ODE.
The solution is implemented with model based design and uses only fundamental linear algebra; consequently, this approach supports automatic code generation for deployment on embedded systems.
The targeting concept was tested via software- and processor-in-the-loop verification on two systems with different processor architectures.
Finally, the method was tested on a laboratory prototype with real measurement data for the monitoring of flexible structures.
The measurement arrangement consists of an embedded system with a chain of $14$ inclinometer sensors connected to it, two additional nodes implement a total of four constraints. 
The problem solved is: the real-time overconstrained reconstruction of a curve from measured gradients. 
Such systems are commonly encountered in the monitoring of structures and/or ground subsidence. 
\end{abstract}

\category{C.3}{Special-Purpose and Application-Based Systems}{Real-Time and Embedded Systems}
\category{G.1.7}{Ordinary Differential Equations}{Boundary value problems}
\category{I.2.2 }{Automatic Programming}{Program transformation}

\terms{Design, Performance, Experimentation}

\keywords{cyber-physical systems, embedded systems, inclinometers,
measurement, numerical solver, ordinary differential equations,
inverse problems, constraints, model based design, automatic code
generation, in-the-loop verification}
\begin{bottomstuff}
Author's address: Chair of Automation, Department Product Engineering, University of Leoben, 8700 Leoben, Austria; URL: automation.unileoben.ac.at; Email: christoph.gugg@unileoben.ac.at;
\end{bottomstuff}
\maketitle
\section{Motivation and Problem Statement}
The original motivation for this work was the development of a
large scale cyber-physical system (CPS) to monitor ground
subsidence and possible deformation of structures during the
construction of the new City-Circle Line subway in Copenhagen,
Denmark. Very stringent geo-mechanical monitoring requirements
have been established for underground construction projects in
urban areas following an accident on March 3, 2009 in Cologne,
Germany: the building of the city's archive collapsed into a
Stadtbahn tunnel under construction on the Severinstra{\ss}e,
killing two people\footnote{An article relating to the incident
can be found at
\url{http://www.ksta.de/html/artikel/1266930835566.shtml}}. The
monitoring concept consists of a large number of vertical holes
sunk along the planned path of the tunnel distributed over a
distance of approximately $15 \unit{km}$. Each of these holes is
equipped with a series of rods, and each rod is equipped with a
pair of inclinometers, effectively forming a chain of
inclinometers. Chains of inclinometers are used in the monitoring
of ground subsidence~\cite{Machan2008} and for measuring the
deformation of structures~\cite{Oleary2012}. Determining the
ground movement from the orientation of the rods is an inverse
problem. Additionally, there are points where constraints are
placed on the construction, for example pillars, which in turn
define initial-, inner- or boundary values for the inverse problems.
Reconstructing the deformation under these circumstances requires
the solution of an inverse boundary value problem for each chain
of rods. Consequently, it is necessary to solve a large number of
inverse initial-, inner- or boundary value problems in real-time for different sets of measurement data. Each
chain of inclinometers is equipped with an embedded system that
acquires and processes the data from the sensors, forming an independent sensor
node.
\revtwo{The individual sensor nodes are part of a larger sensor network. Decentralized processing of measurement data introduces an implicit form of parallelism thanks to distributed computing. The network's bandwidth demands are lowered due to the higher information density.}

Necsulescu also identified the necessity of solving inverse
problems in critical infrastructure
monitoring~\cite{Necsulescu2005}. Lee~\cite{Lee2012} identified
that  predictable real-time solutions of complex systems, with an
understandable concurrency, are a key issue for future
developments of CPS. He points out that this issue was
inadequately dealt with in the past. There are numerous
engineering and scientific applications which require the
real-time solution of inverse problems, e.g.~\cite{Loh1996}. Therefore, this is clearly an area of research which is of significance.

\section{Scope of the Article}
This article develops a new method for the numerical solution of inverse problems based
on a matrix algebraic approach. It provides global least squares solutions to inverse
initial-, inner- or boundary value problems. The method has been
developed specifically with the aim of solving inverse problems
associated with measurement systems in an efficient manner, whereby multiple measurements
are performed over time and repeated solutions of the same equation are required. \revtwo{The goal is to directly embed the solver onto the sensor node's hardware.} The main
contributions of the article are:
\begin{enumerate}
    \item A new algebraic approach to the numerical solution of inverse problems is
    derived. The method splits the calculations into two portions:
    a preparatory (offline) computation and a
    run-time (online) computation. The run-time computation is repeatedly performed with each
    new measurement. Solving the inverse problem at run-time is
    reduced to one matrix multiplication and one vector addition.
    In this manner, the exact number of floating point operations (FLOPs)
    is known a-priori, $W(n) = 2 \, n^2$, where $n$ is the number
    of measurement points. Additionally, the memory requirements are
    known in advance. Consequently, a strict \revtwo{upper-bound $\mathcal{O}(n^2)$} can
    be determined for the execution time on a given processor. This makes the method, by definition, suitable for
    real-time applications. Furthermore, the covariance
    propagation for perturbations of the sensor inputs to
    the solution is derived. This enables
    the computation of a confidence interval for the solution. The
    run-time computational complexity of estimating the
    confidence is $\mathcal{O}(n)$.

    Extensive \revfour{model-in-the-loop (MIL)} testing of the method on a personal computer (PC) is presented to validate
    the method. The results of a classical Runge-Kutta type approach are compared
    with those obtained using the new approach. The results demonstrate the accuracy
    of the method and its numerical efficiency.

    \item \revtwo{A model based design (MBD) approach is presented which
    enables the system formulation at an abstract level. The presented model
    only utilizes fundamental linear algebra operations such as matrix multiplication
    and vector addition; consequently, automatic generation of C code becomes possible.
    Software-in-the-loop (SIL) verification is used to proof the functional equivalence
    of the model and the generated code. Embedded targeting enables the deployment of
    the code directly onto a microcontroller. The results computed by the embedded
    processor are compared to the results computed by the model running on a PC via
    processor-in-the-loop (PIL) verification. The viability of the model is demonstrated
    on a very limited, yet cheap and available, $8$-bit microcontroller.} \revfour{Furthermore,
    a laboratory setup with a chain of inclinometers  mounted on a flexible structure
    demonstrates the applicability of the model for real measurement data.}
\end{enumerate}
\section{Continuous Measurement Model} \label{sec:measurementModel}
%
The measurement model is central to this article: it defines the
class of problem which is being solved. Furthermore, it defines
the requirements for the MBD environment. The aim is to use MBD to
automatically generate the functionally equivalent code which is capable of solving any
example of this problem on an embedded system in
real-time.

The class of inverse problems being considered in this article
consist of an ordinary differential equation (ODE) of degree $m$
of the form
\begin{equation}
    a_m(x) \, y^{(m)} + a_{m-1}(x) \, y^{(m-1)} + \ldots
    + a_1(x) \, y' + a_0(x) \, y = g(x)\label{eqn:de1},
\end{equation}
where $y$ is a function of $x$, $y^{(i)}$ is the notation for the
$i^{\text{th}}$ derivative of $y$ with respect to $x$, $a_i(x)$
are the coefficient functions and $g(x)$ is the forcing function.
Additionally, a minimum of $m$ independent initial-, inner- or
boundary values are required to ensure that there is a unique
solution to the equation. The $n$ measurements, forming the vector
$\V{g}$, correspond to discrete samples of the forcing function
$g(x)$. The $n$ measurements may emanate from $n$ sensors forming
a spatial array or from a time sequence of $n$ measurements from
one single sensor. In this class of problems, the forcing function $g(x)$,
the input, is considered to be perturbed, since it is formed from
measurements which are subject to noise. Only the forcing function
$g(x)$ changes from one measurement to the next. The task is to
recompute $y(x)$ for each new measurement $g(x)$. This type of
problem occurs, for example, in the monitoring of
structures~\cite{Burdet2002,Golser2010,harker2013}.

The new method can, however, deal with overconstrained systems,
i.e., there are $p$ independent constraints whereby $p > m$. The
initial-, inner- or boundary values correspond to constraints on
the function value $y(x)$ or its derivatives $y^{(i)}(x)$ at
specific $x$ locations. Both Dirichlet and Neumann boundary
conditions are special cases of such constraints. The nature of
the constraints determines if the system is considered to be an
initial value (IVP) or boundary value (BVP) or inner value
problem.

One peculiarity of this class of inverse problems is: that the
abscissae, i.e., the positions where the solutions are required,
is determined by the measurements; these positions are called the
nodes. In the case of a chain of sensors, the physical position of
the sensor corresponds to the abscissae $x$. In temporal sequences,
it is the time points of the individual measurements which define
the abscissae. \revfive{Consequently, we are not free to select the
positions of where the ODE is to be solved.} This precludes the use of
variable step size algorithms. A further consequence is that a general
framework for this type of inverse problem has to be capable of
computing the solution for arbitrary nodes.
\section{Theory of Ordinary Differential Equations}\label{sec:PreTheory}
Some preliminary theory is required if an objective evaluation of
previous work is to be performed. The numerical solution of an
inverse problem requires the discrete approximation of a continuous system. Consequently, we can derive
properties of the continuous operations which must be fulfilled by
the corresponding discrete operators.
We first define the continuous domain differential operator $D$
such that, $D^{(i)} \, y \equiv y^{(i)}$. Most commonly, the
discrete implementation of the differentiating matrix is
implemented using polynomial interpolation. The properties of $D$
with respect to a polynomial are essential to the desired behavior
of numerical differentiation. Defining a power series
approximation for $y$ with coefficients $c_i$,
\begin{equation}
   y = \sum_{i=0}^{m} c_i \, x^i .
\end{equation}
Applying the differential operator $D$ yields,
\begin{equation}
   y^{\prime} = D \, y = \sum _{i=0}^{m} i \, c_{{i}} \, {x}^{i-1}
   = \sum _{i=1}^{m} i \, c_{{i}} \, {x}^{i-1}.
\end{equation}
By definition of the derivative, the
constant portion of the polynomial differentiates to
zero, hence the constant coefficient $c_0$ vanishes.
\revfive{ We assume that $D$ is composed of formulae which are
consistent, in the sense that in the limit they define a
derivative. If this is the case then the matrix $\M{D}$ should
satisfy the following properties, such that $\M{D}$ is a
consistent discrete approximation to the continuous operator $D$:}
\begin{enumerate}
   \item The matrix $\M{D}$ must be rank-$1$ deficient; i.e,
   its null space is of dimension one.
   \item The null space of $\M{D}$ must be spanned by the constant
   vector $\VO \,
   \alpha$; equivalently, the row-sums of $\M{D}$ are all zero,
   \begin{equation}
       \M{D} \, \VO \, \alpha = \VZ.
   \end{equation}
\end{enumerate}
%
%
These conditions ensure that the differentiating matrix $\M{D}$ is
consistent with the continuous domain definition of the
derivative. Given that, interpolating polynomials are unique, the
formula for the derivative should be independent of the particular
polynomials chosen for interpolation. However, differences do lie
in the numerical behavior of different formulas; regardless, a
given set of nodes, $\V{x}$, should uniquely define the
differentiating matrix of a given polynomial degree of accuracy.

For the purpose of treating ODEs, we
use the general notion of a linear differential
operator~\cite{Lanczos1997}. Specifically, by substituting the
continuous differential operator $D$ for the differentials
$y^{(i)}$ in Eqn.~(\ref{eqn:de1}) yields,
\begin{equation}\label{eqn:de2}
   a_m(x)  \, D^{(m)}\, y + a_{m-1}(x) \, D^{(m-1)} \, y + \ldots
   + a_1(x) \, D\,y + a_0(x) \, y = g(x).
\end{equation}
Factoring $y$ to the right yields,
\begin{equation}\label{eqn:de3}
   \left\{a_m(x) \, D^{(m)} + a_{m-1}(x) \, D^{(m-1)} + \ldots
    + a_1(x) \, D + a_0(x) \right\} \, y = g(x).
\end{equation}
The linear differential operator $L$ for the continuous equation can now be defined as,
\begin{equation}\label{eqn:de4}
   L \defas a_m(x) \, D^{(m)}+ a_{m-1}(x) \, D^{(m-1)} + \ldots
   + a_1(x) \, D + a_0(x).
\end{equation}
Consequently, Eqn.~(\ref{eqn:de1}) is written as,
\begin{equation}
   \boxed{L \, y = g(x).}
\end{equation}
\section{Overview of Numerical ODE Solvers}\label{sec:previous}
The \textit{Taylor matrix} uses the known analytical relationship between the
coefficients, $\V{s}$, of a Taylor polynomial and those of its
derivatives, $\dot{\V{s}}$, to compute a differentiating matrix
$\M{D}$ for the solution of ODEs~\cite{Kurt2008}. The matrix $\M{D}$ together with the matrix of basis
functions arranged as the columns of the matrix $\M{B}$ are used
to compute numerical solutions to the differential equations. The
method of the Taylor matrix was extended to the computation of
fractional derivatives~\cite{Keskyn2011}. The most serious problem
associated with the Taylor matrix approach is that it requires the
inversion of the Vandermonde matrix, a process which is
numerically unstable. The errors in the differentiating matrix are
strongly dependent on the degree of the polynomial, i.e., the
number of nodes and the node placement.

A \textit{Chebyshev matrix} approach was presented by Sezer
\cite{Sezer1996} and
others~\cite{Welfert1997,Weideman2000,Driscoll2008,Jewell2013}.
The approach is fundamentally the same as for the Taylor matrix,
whereby the Chebyshev polynomials are used as an alternative to
geometric polynomials. The advantage of defining polynomials on
the Chebyshev points is that they deliver stable polynomials and
differentials. The main disadvantage, however, is that the
numerical solution to the differential equations is restricted to
the locations of the Chebyshev points; this lacks the generality
needed for inverse problems\footnote{This is not dismissing the
Chebyshev methods, it simply points out that they are limited in
their applications. Furthermore, the methods in this article work
for truly arbitrary node placements. Consequently, the Chebyshev
polynomials are only a special case.} being considered here.

Synthesizing differentiating matrices for arbitrary nodes
is an issue one might assume has been sufficiently dealt with in
literature. However, a closer examination of literature and
textbooks shows that some clarification is still necessary. Most
books on \textit{spectral and pseudo-spectral techniques},
e.g.,~\cite{Fornberg1998}, approach differentiation matrices from
the view point of simulation and do not consider the connotations
of inverse problems. In a simulation, it is in general possible to
select the position of the nodes, so that they are well
suited to the solution method, e.g., it is possible to use either
the Chebyshev or Legendre collocation nodes. This luxury is not
given with inverse problems; the placement of the sensors may be
arbitrary and or the time points for which solutions are required
are evenly spaced. Consequently, it is necessary to generate
differentiating matrices for truly arbitrary nodes.

There are a number of papers~\cite{Welfert1997,Weideman2000} which
explicitly claim to compute \textit{differentiating matrices using
global methods for arbitrary nodes} and there are some toolboxes
which suggest this is possible~\cite{Jewell2013}. The published
code for all these methods generate degenerate differentiating
matrices with null spaces of dimensions higher than one. That is,
they do not fulfill the prerequisites defined in
Section~\ref{sec:PreTheory}. In contrast, the \textit{local
polynomial approximation to differentiation}~\cite{Savitzky1964}
with correct end-point formulas~\cite{burden2005} generates a
consistent matrix. The poor behavior of high order polynomial
interpolation and differentiation is due to Runge's phenomenon,
which will be always be present due to the uniqueness of
interpolating polynomials; hence, approximations of relatively low
degree are preferable to global approaches. The published
methods~\cite{Welfert1997,Weideman2000} work reliably only for
very small problems\footnote{This can be verified by running the
available code with $n = 20$. Testing the resulting $\M{D}$ matrix
or its singular values reveals that a null space of higher
dimension is present. As a consequence, the matrix does not
fulfill the necessary prerequisites.}, $n \leq 10$; this is not
sufficient to address most real inverse problems encountered in
engineering applications. We conclude that global techniques for
computing differentiating matrices are not applicable to large
scale inverse problems.

\textit{Finite difference methods}~\cite{strikwerda200,smith1985} rarely
deal with higher degree approximations, typically $3$ or $5$ point
formulas are used. The issue of correct end point formulas is
sacrificed for the advantage of band diagonal matrices. In general, these techniques deal with
Dirichlet and possibly Neumann boundary conditions. However, they
provide no method of implementing general boundary conditions of the form
\begin{equation}
    D^{(i)} \, y(x_j) = d,
\end{equation}
where $D^{(i)}$ represents the $i^{\text{th}}$ derivation of $y$ evaluated at the point $x=x_j$ with the value $d$. There may be $p \geq m$ such constraints.

A \textit{new matrix approach for the solution of inverse
problems}, associated with monitoring of structures using
inclinometers, was presented~\cite{Oleary2012} and generalized
in~\cite{harker2013}. It was proven that ODEs can be formulated as
a least squares problem with linear constraints, of the form:
    \begin{equation}\label{eqn:start}
        \M{L} \, \V{y} = \V{g} \eqtext{subject to}
        \MT{C} \, \V{y} = \V{d},
    \end{equation}
whereby $\M{L}$ is the discretized linear differential operator,
$\V{y}$ is the solution vector sought (function values), $\V{g}$ is the discrete
forcing function (measurement values), $\M{C}$ defines the type of constraints and
$\V{d}$ are the values of the constraints. The least squares
solution makes the method suitable for problems where the forcing
function $g(x)$ is perturbed.

%
%
%
%

The continuous linear differential operator $L$ in Eqn.
(\ref{eqn:de4}) is discretized as the matrix $\M{L}$, such that
\begin{equation}
    \M{L} \defas \M{A}_m \, \M{D}_m + \M{A}_{m-1} \, \M{D}_{m-1} + \ldots +
     \M{A}_1 \, \M{D} + \M{A}_0,
\end{equation}
where $\M{A}_i = \mathrm{diag}(a_i(\V{x}))$, the matrix $\M{D}_i$
is a local discrete approximation with support length $l_s$ to the
continuous differential operator $D^{(i)}$. Care is taken to implement the
correct end-point formulas, ensuring the degree of approximation
is constant for the complete support. The details of generating
these matrices can be found in~\cite{harker2013}, as can the
explanation for the generation of the constraints $\MT{C}\, \V{y} = \V{d}$.
Furthermore, MATLAB toolboxes are
available~\cite{harker2013DOP,harker2013ODE} for all the functions
required in this article.
\section{Solving the Inverse Problem}
Previously the problem in Eqn.~(\ref{eqn:start}) was solved using
an efficient and accurate solution which is found in~\cite[Chapter
12]{Golub}. In this paper we take a different approach to
partitioning the numerical computations, which takes advantage of
the fact that the inverse problem is to be solved repeatedly.
Fundamentally, the new approach delivers exactly the same explicit
solution; however, through the new partitioning of the computation
it is possible to ensure that the numerical work $W(n)$ and the
memory required are run-time are known exactly in advance.
Consequently, an exact upper-bound for the execution time can be
determined, this by definition makes the solution suitable for
real-time applications\footnote{A \textit{real-time system} is defined as any information processing activity or system which has to respond to externally generated input stimuli within a finite and specified period~\cite{young1982}.}.

The computation of the solution is separated into two portions:
\begin{enumerate}
    \item The preparatory computations which can be performed
    offline. They are characteristic for the equation being
    solved and change neither with the acquisition of new
    measurement data, nor with new values for the boundary
    conditions. These computations need not be performed on the
    embedded system and may be computed with higher precision
    arithmetic on a host system if necessary.
    \item The online computation, which must be performed
    repeatedly with each new set of sensor data. This is the
    solution which is computed explicitly on the embedded system in
    real-time.
\end{enumerate}

\subsection{Preparatory Computations}

\begin{figure}
  \begin{minipage}[t]{0.49\columnwidth}
    \centering
    \includegraphics[width=65mm]{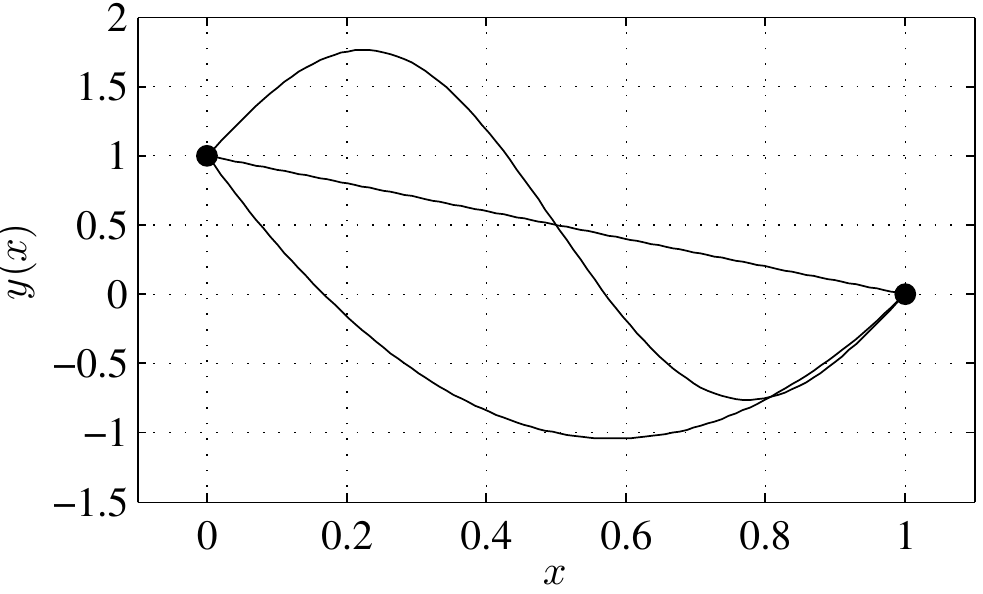}
    \caption{Three different solutions to the constraints $y(0) = 1$ and $y(1) = 0$.}
    \label{fig:ConstParticular}
  \end{minipage}
  \hfill
  \begin{minipage}[t]{0.49\columnwidth}
    \centering
    \includegraphics[width=65mm]{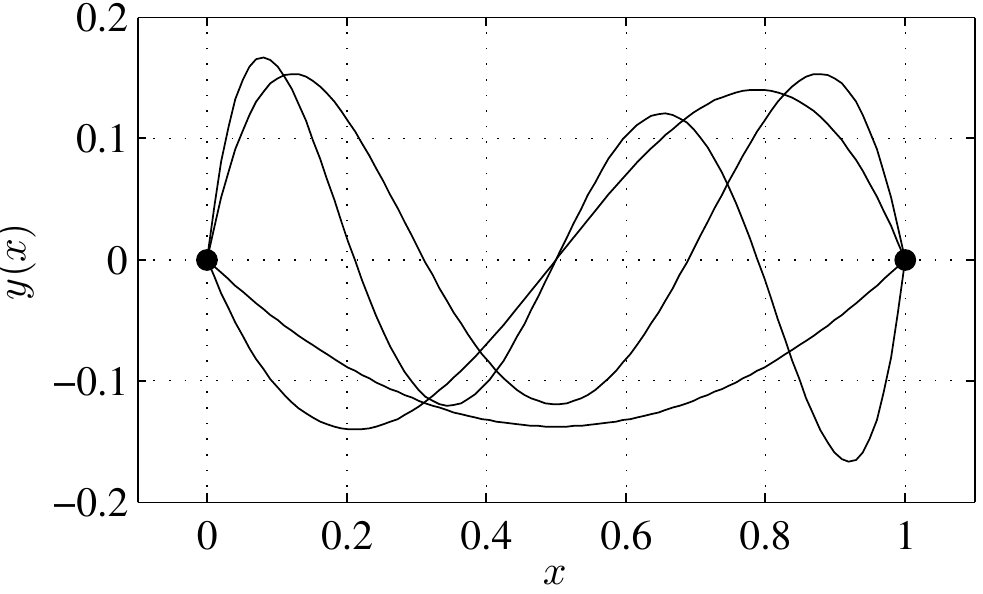}
    \caption{The first four discrete orthogonal constrained polynomials for the
    constraints $y(0) = 1$ and $y(1) = 0$, i.e., the first $4$ basis functions from $\M{F}$.}
    \label{fig:ConstAddmiss}
  \end{minipage}
\end{figure}

The constraints on the solution are defined by,
\begin{equation}\label{eqn:const}
    \MT{C} \, \V{y} = \V{d}.
\end{equation}
Each column of $\M{C}$, together with the corresponding row of
$\V{d}$, defines a constraint. Consequently, $p = \rank{\M{C}}$ is
the number of linearly independent constraints. Additionally, the
constraints must be consistent, i.e., $\V{d} \in \range{\MT{C}}$. A
minimum of $p \geq m$ constraints are required to ensure a unique
solution to an ODE of degree $m$.
We now define the matrices: $\M{P}$, such that $\range{\M{P}} =
\range{\M{C}}$, i.e., $\M{P}$ is the Moore-Penrose
pseudo inverse of $\MT{C}$, hence $\M{P} = \{\MT{C}\}^{+}$; $\M{F}$, an orthonormal basis function
set for the null-space of $\MT{C}$, i.e., $\MT{F}\,\M{F} = \M{I}$
and $\range{\M{F}} = \nullS{\MT{C}}$; and $\M{H} \defas \M{P} +
\M{F} \, \M{R}$, where $\M{R}$ is an arbitrary matrix. In this
manner the solution for $\V{y}$ can be parameterized as,
\begin{equation}\label{eqn:nonUnique}
     \V{y} = \M{H} \,\V{d} + \M{F} \, \V{\beta},
\end{equation}
where $\V{\beta}$ is the parameter vector. It is important to realize that neither $\M{H}$ nor $\M{F}$ are
unique. Any function which fulfills the constraints is a valid
selection for $\V{y}_c$. A function $\V{y}_c$ which fulfills the
constraints can be defined as,
\begin{align}
    \V{y}_c & \defas \M{H} \,\V{d},\\
    & = \left\{ \M{P} + \M{F}\,\M{R} \right\} \, \V{d}.
\end{align}
The matrix $\M{R}$ is arbitrary, consequently the values can be
selected so that $\V{y}_c$ fulfills additional conditions without
altering the solution for $\V{y}$. In
Fig.~\ref{fig:ConstParticular} three different solutions for the
constraints $y(0) = 1$ and $y(1) = 0$ are shown, to demonstrate
this fact. It may be advantageous for a specific problem to select
a particular solution for $\V{y}_c$ which has desirable
properties; for example, when solving the ODE for a cantilever it
may be appropriate to select a polynomial solution for $\V{y}_c$,
since the solution to the ODE is known to be a polynomial. More
formally: the matrix $\M{H} = \left\{\MT{C}\right\}^{-}$ is a
generalized inverse~\cite{ben2003} of $\MT{C}$. A generalized
inverse $\M{A}^{-}$ of a matrix $\M{A}$ fulfills the condition,
\begin{equation}
    \M{A} \, \M{A}^{-} \, \M{A} = \M{A}.
\end{equation}
The Moore-Penrose pseudo inverse is the particular generalized
inverse, where $\M{R} = \M{0}$; it yields an inverse which
minimizes the $2$-norm of the solution vector; alternatively, a QR
decomposition can be used to compute a generalized inverse which
leads to a solution vector with a minimum number of nonzero
entries. The selection of an appropriate solution for $\V{y}_c$ is
more important when solving inverse problems, since it has
implications for the implementation of regularization.


The orthonormal basis functions $\M{F}$ for the null-space of
$\MT{C}$ are also not unique. They can be obtained directly from
$\MT{C}$ by applying QR decomposition and partitioning Q according
to the $\rank{\M{R}}$. Alternatively, constrained basis functions,
e.g. constrained polynomials, can be used to implement a set of
orthogonal basis functions $\M{F}$, Fig.~\ref{fig:ConstAddmiss}
shows an example of such admissible functions for the constraints
$y(0) = 1$, and $y(1) = 0$. In the case of inverse problems
constrained basis functions offer a method of implementing
spectral regularization~\cite{Oleary2012}. The MATLAB library
required to generate discrete orthogonal constrained polynomials
is available at~\cite{harker2013DOP}.

\begin{equation}
    \MT{C} \, \M{F} = \M{0} \eqtext{and hence} \MT{C} \, \M{F} \, \V{\beta} = \VZ.
\end{equation}
%
%
%
Substituting Eqn.~(\ref{eqn:nonUnique}) for $\V{y}$ in
$\M{L}\,\V{y} = \V{g}$ now yields an unconstrained algebraic
equation for the ODE,
\begin{equation}\label{eqn:unConstrained}
    \M{L}\, \left\{ \M{H} \, \V{d} + \M{F} \, \V{\beta} \right\} =\V{g}.
\end{equation}

In the class of inverse problems being considered in this article,
the forcing function $\V{g}$ is formed from the measurement values
which are perturbed, i.e., is subject to noise. Consequently, the
solution of Eqn.~(\ref{eqn:unConstrained}) is formulated as a
least squares problem to obtain the unique global minimum of
\begin{equation}\label{eqn:LSEqn}
    \min_{\V{\beta}}{\| \M{L}\,\M{H} \, \V{d} +  \M{L} \, \M{F} \, \V{\beta} - \V{g}
    \|^2_2}.
\end{equation}
The least squares approach has been selected since it delivers a
maximum likelihood solution in the case that $\V{g}$ is perturbed
by Gaussian noise. A further advantage of the global least squares
formulation is that the solution has no implicit direction of
integration. Avoiding a direction of integration eliminates the
problem of accumulation of errors, as are typical with IVP
approaches such as Runge-Kutta. Additionally, the least squares
approach yields a solution which is globally minimum with respect
to all errors in Eqn.~(\ref{eqn:LSEqn}), i.e., it is also
minimizing the consequences of the errors in the numerical
computations. Consequently, the method is suitable for solving
both perturbed and unperturbed problems. Now solving the
minimization problem defined by Eqn.~(\ref{eqn:LSEqn}) yields,
\begin{equation}
     \V{\beta} =
    \left\{ \M{L} \, \M{F} \right\}^+ \, \left\{ \V{g} - \M{L}\,\M{H} \, \V{d} \right\}
    + \M{K} \, \V{\gamma},
\end{equation}
where $\M{K}$ is an orthonormal vector basis set for the
null-space of $\M{L} \, \M{F}$, i.e., $\MT{K} \, \M{K} = \M{I}$
and $\spanS{\M{K}} = \nullS{\M{L}\,\M{F}}$. This equation is now
expanded into three relevant terms,
\begin{align}
     \V{\beta} & = \left\{ \M{L} \, \M{F} \right\}^+ \, \V{g}
     - \left\{ \M{L} \, \M{F} \right\}^+ \, \M{L}\,\M{H} \, \V{d}
     + \M{K} \, \V{\gamma}.
\end{align}
A non-empty vector basis set $\M{K}$ indicates that the linear
differential operator $\M{L}$ is not sufficiently constrained to
ensure a unique solution, i.e., there is no unique solution to the
problem being posed. The requirement for a unique solution is
\begin{equation}
    \mathrm{rank}
    \begin{bmatrix}
        \M{L} \\
        \MT{C}
    \end{bmatrix}
    = n,
\end{equation}
where $n$ is the number of nodes. This is a method of determining
if the problem is well defined. Alternately, the singular values
of the matrix can be used to determine if the problem is
numerically well posed. We will now assume that the problem is
well posed: with this, the term involving $\V{\gamma}$ vanishes.
Now back-substituting for $\V{\beta}$ in
Eqn.~(\ref{eqn:nonUnique}), yields,
\begin{align}\label{eqn:sol}
\V{y}
     &= \M{F} \, \left\{ \M{L} \, \M{F} \right\}^+ \, \V{g} 
     + \M{H} \, \V{d} - \M{F} \,\left\{ \M{L} \, \M{F} \right\}^+ \,
     \M{L}\,\M{H} \, \V{d}.
\end{align}
Defining the following abbreviations:
\begin{equation}
    \M{M} \defas \M{F} \, \left\{ \M{L} \, \M{F} \right\}^+
    \hspace{1cm}
    \text{and}
    \hspace{1cm}
    \M{N} \defas \left\{\M{I} - \M{M} \, \M{L}\right\} \, \M{H},
\end{equation}
yields,
\begin{align}
    \V{y} &= \M{N} \, \V{d} + \M{M} \, \V{g} \\
     &= \V{y}_h + \V{y}_p.
\end{align}
The homogeneous portion of the solution $\V{y}_h = \M{N} \, \V{d}$
is only dependent of the constraint values and the particular
solution $\V{y}_p = \M{M} \, \V{g}$ is only dependent on the
forcing function, i.e., the measurement values. In the problems
considered in this paper the constraint values do not change from
one measurement to the next. Consequently, $\V{y}_h$ can be
computed a-priori and made available as a vector of constraints
for the run-time computation.
\subsection{Run-Time Computation}
Both $\M{M}$ and $\V{y}_h$ are computed a-priori. A standard
PC with higher precision arithmetic can be used for
these computations. In this manner, the final errors in $\M{M}$
and $\V{y}_h$ are dominated by the rounding effects of converting
the double precision values to single precision for the embedded computation, should
the embedded system not support double precision arithmetic.
Substituting $\M{M}$ and $\V{y}_h$ into Eqn.~(\ref{eqn:sol})
yields,
\begin{equation}\label{eqn:TheBigOne}
\boxed{
    \V{y} = \M{M} \, \V{g} + \V{y}_h.
    }
\end{equation}
Only the vector of sensor data $\V{g}$ changes with each
measurement. Consequently, the run-time solution of the inverse
problem is reduced to a single matrix multiplication and a vector
addition. This makes the repeated computation of the solution very
efficient. Given $n$ measurement values, the computational cost $W(n)$ is,
\begin{equation}\label{eqn:big}
    W(n) = 2 n^2.
\end{equation}
For example, a sensor chain with $n = 21$ inclinometers would
require $W(n) = 882$ FLOPs to solve the inverse problem. The
computation effort reduces to
\begin{equation}\label{eqn:big2}
    W(n) = n^2.
\end{equation}
if the processor architecture being used supports a
multiply-accumulate\footnote{\revfour{See for example the specifications
for the ARM Cortex Microcontroller Software Interface Standard (CMSIS) at \url{http://www.arm.com/products/processors/cortex-m/}.}} operation. Both the exact number
of FLOPs and memory required are known prior to the run-time
computation. This enables the computation of a strict upper bound
for the execution time of the equation. Consequently, the method
is, by definition, suitable for
real-time applications. The computational complexity,
$\mathcal{O}(n^2)$, is independent of the placement of the nodes,
the equation being solved and the support length selected.
%
%
%
%
\subsection{Error Estimation and Confidence Interval}
There is uncertainty associated with the solution of any inverse
problem. Regularization is used to control this
uncertainty. The aim now is to quantify the uncertainty associated
with the solution presented here. The are two primary
sources of possible errors involved in the computation of $\V{y}$ in
Eqn.~(\ref{eqn:TheBigOne}):
\begin{enumerate}
    \item Errors in the values contained in $\M{M}$ and $\V{y}_h$.
    The numerical testing (see Section~\ref{secTesting}) demonstrates
    that these errors are negligible, when computed in double precision,
    in comparison to realistic perturbations of $\V{g}$. For some
    embedded systems it is necessary to reduce the values from
    double to single precision. Software-in-the-loop (SIL)
    and processor-in-the-loop (PIL) testing are used to quantify
    these errors. In Section~\ref{sec:exptest} it is experimentally
    verified that these errors can be ignored.
    \item The errors at run-time are dominated by the perturbations
    of $\V{g}$, these errors are orders of magnitude larger than
    the residual numerical errors in $\M{M}$ and $\V{y}_h$. Consequently,
    only errors in $\V{g}$ are considered for the covariance propagation.
   \revfive{There is also an approximation error in $\M{M}$ based on the choice of the interpolating functions. These may not be insignificant depending on the nature of the solution $\V{y}$.}
\end{enumerate}

The following computation assumes that only the forcing function
$\V{g}$ is subject to Gaussian perturbation. The covariance
$\M{\Lambda}_{\V{y}}$ associated with the computation of
$\V{y}$ using Eqn.~(\ref{eqn:TheBigOne}), can be
explicitly~\cite{Brandt1998} calculated as
\begin{equation}\label{eqn:stats1}
    \M{\Lambda}_{\V{y}} = \M{M} \, \M{\Lambda}_{\V{g}} \, \MT{M},
\end{equation}
where $\M{\Lambda}_{\V{g}}$ is the covariance of the forcing function.
In practical applications we determine the magnitude of the noise
component for each sensor, using dedicated noise measurements. In
this case, and assuming that the noise is independent identically
distributed (i.i.d.) Gaussian noise with standard deviation
$\sigma_{\V{g}}$, then,
\begin{equation}\label{eqn:stats2}
    \M{\Lambda}_{\V{g}} = \sigma_{\V{g}}^2 \, \M{I},
\end{equation}
where $\sigma_{\V{g}}$ is a measured value. Substituting this into
Eqn.~(\ref{eqn:stats1}) yields,
\begin{align}
        \M{\Lambda}_{\V{y}} = \sigma_{\V{g}}^2 \, \M{M} \, \MT{M}.
\end{align}
An upper-bound estimate \revfive{within a given confidence interval} for the vector of standard deviations for
$\V{y}$, is computed as,
\begin{equation}\label{eqn:statsFinal}
    \boxed{
        \V{\sigma}_{\V{y}} = \sigma_{\V{g}} \, \V{s},
        }
\end{equation}
\revfive{
where $\M{M} = (m_{ij})$ and the individual elements of the unscaled standard deviation $\V{s}$ are $s_i = (\sum_{j=1}^n m_{ij}^2)^{1/2}$, i.e. the square root of the diagonal elements of $(\M{M} \, \MT{M})$.}
%
%
This term can be computed a-priori; consequently, the run-time computational
complexity for determining the standard deviation of each solution
point is $\mathcal{O}(n)$.

Alternatively, the error vector $\V{\epsilon}$ can be computed for
each measurement as the difference between the forward and inverse
problem, i.e.,
\begin{equation}\label{eqn:errorEst}
        \V{\epsilon} = \V{g} - \M{L} \left(\M{M} \, \V{g} +
        \V{y}_h\right).
\end{equation}
A Kolmogorov-Smirnov test can be applied to $\V{\epsilon}$  to
determine if it is Gaussian. This yields additional information on
the suitability of the model for the specific measurement. Given
the standard deviation, the confidence interval with a specific
degree of certainty is computable via the inverse
Student-\textit{t} distribution~\cite{Brandt1998}.
%
%
%
%
%
\section{Model-in-the-Loop Testing}\label{secTesting}
The aim of this section is to verify the numerical accuracy and
efficiency of the new method on a PC, embedded testing
is presented later. We have chosen to solve unperturbed problems
for the first tests, since these enable the comparison with
analytical solutions and with standard engineering approaches such
as Runge-Kutta methods\footnote{The MATLAB {\tt ode45} implementation of a
Runge-Kutta method was used for this purpose.}. The unperturbed
tests enable the separation of the errors involved in computing
$\M{M}$ and $\V{y}_h$ from those resulting from the perturbation
of $\V{g}$.
It is difficult to define a truly objective method of comparing
solution approaches which are fundamentally different\footnote{To
support independent verification of our results, we have made the
MATLAB code available which we used to generate all the results
presented in this section, see \url{http://www.mathworks.com/matlabcentral/fileexchange/45947}.}. Each
approach has its own weaknesses and strengths. The tests have been
devised to reflect, as close as possible, the conditions which are
to be expected from the application specific method.
\subsection{Test A: Initial Value Problem 1} \label{secIVP1}
The ODE (details can be found in~\cite{Adams}) is a third
order ($m=3$) non-homogeneous ODE with constant coefficients $a_i$ and $p=3$ constraints. The equation is
\begin{align}\label{eqn:IVP3}
    & y^{(3)} + 3\,y'' + 3 y' + y = 30 \, \me^{-x} \eqtext{given}\\
    &
    y(0) = 3,
    \hspace{5mm}
    y'(0) = -3,
    \hspace{5mm}
    y''(0) = - 47 \notag
\end{align}
in the interval $0 \leq x \leq 8$. The analytical solution to this
equation is
\begin{equation} \label{eqn:testAanalytical}
    y(x) = (3 - 25\,x^2 + 5\,x^3)\,\me^{-x}.
\end{equation}
In the case of the inverse problems being addressed, the position
of the solution points is determined by the measurement. To
simulate this condition, the {\tt ode45} solver~\cite{Shampine1997}
in MATLAB has been used to solve this differential equation. \revtwo{This
is a variable step size method which yields both a vector of
abscissae $\V{x}$ consisting of $n=77$ points and the solution vector $\V{y}$. Exactly those $n=77$ points on the abscissae and a support length $l_s = 9$ was used for the test of the new method.} Using the {\tt ode45} solver in addition to the analytical solution enables the comparison of the new method with well established techniques.
\begin{figure}
  \begin{minipage}[t]{0.49\columnwidth}
    \centering
    \includegraphics[width=65mm]{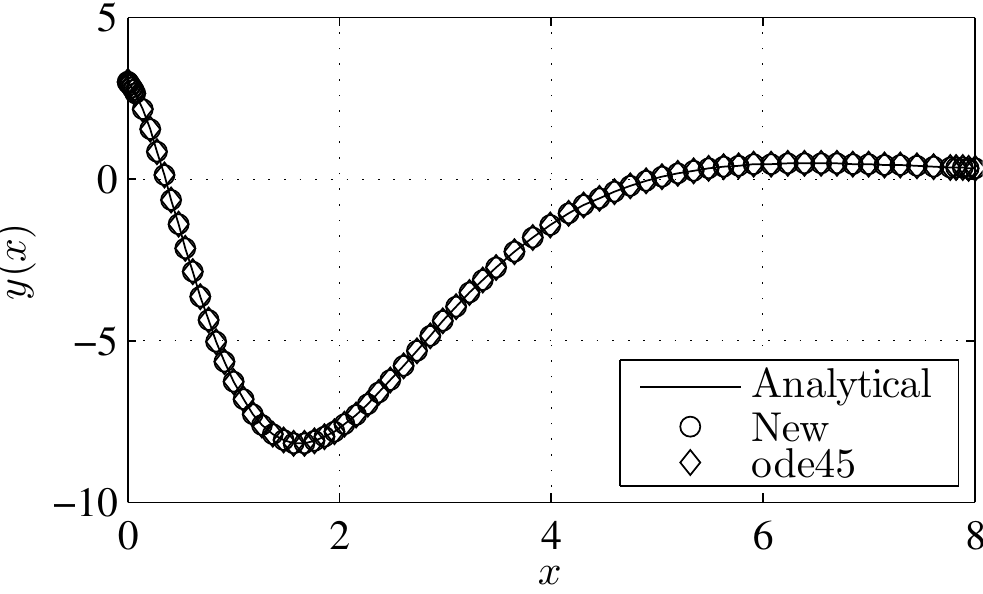}
    \caption{\revtwo{The plot shows the analytic solution from Eqn.~(\ref{eqn:testAanalytical}), the solution with the new method with $l_s=9$ as well as the solution by a Runge-Kutta {\tt ode45} method. All solutions are evaluated at exactly the $n=77$ points provided by the {\tt ode45} method.}}
    \label{fig:ivpExp3a}
  \end{minipage}
  \hfill
  \begin{minipage}[t]{0.49\columnwidth}
    \centering
    \includegraphics[width=65mm]{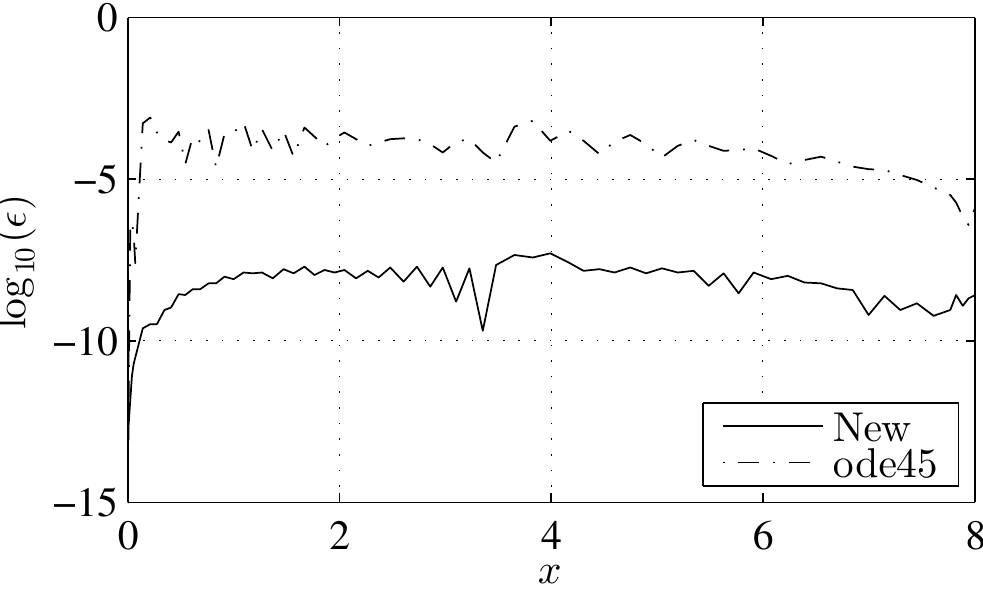}
    \caption{\revtwo{Residual errors: difference of the new numerical solution vs. the analytical solution and the solution by a Runge-Kutta {\tt ode45} method vs. the analytical solution. The new method is approximately $4$-orders of magnitude more accurate than the {\tt ode45} method, for exactly the same abscissae.}}
    \label{fig:ivpExp3b}
  \end{minipage}
\end{figure}
The results of the three computations are shown in
Fig.~\ref{fig:ivpExp3a}. The residual errors, i.e., the difference
between the analytical solution, the new method and the {\tt
ode45} solutions are shown in Fig.~\ref{fig:ivpExp3b}. The
$2$-norm of the residual errors $|\V{\epsilon}|_2$ and the computation
time for $k = 10000$ iterations\footnote{It is not the absolute
times which are important, since they will change from one
platform to another. It is the relative speed which shows the
potential performance of the new method.} for the solution of the ODE are given in Table~\ref{tab:test1} for the {\tt
ode45} method as well as the new method.
\begin{table}[ht]
  \centering
  \caption{The $2$-norm of the residual errors $|\V{\epsilon}|_2$
  and the computation time for $k = 10000$ iterations for the solution of the ODE,
  for the {\tt ode45} method and new method (New). These computations were
  performed with an Intel Core $2$ Duo CPU P8600 at $2.4 \unit{GHz}$ with $2.9 \unit{GB}$ RAM.}
  \label{tab:test1}
  \begin{tabular}{l|c|c}
    method & $|\V{\epsilon}|_2$ & time ($k=10000$)\\
    \toprule
    {\tt ode45} & $1.79 \, 10^{-3}$ & $29.823728 \unit{s}$  \\
    New & $1.14 \, 10^{-7}$ & $0.061681 \unit{s}$\\
    \midrule
    \bottomrule
  \end{tabular}
\end{table}
The first observation is that the residual numerical errors for
the new method are approximately $4$-orders of magnitude smaller than
with the {\tt ode45} method and the computation is almost $500$
times faster. Reducing the error bound for the Runge-Kutta method will improve the numerical accuracy, but at the expense of computational effort.
\revfive{
In order to reduce the error by $4$-orders of magnitude in the Runge-Kutta solution, an even higher degree Runge-Kutta method must be formulated, which in turn would require unreasonably high computational effort.
}
The very small errors are significant: since, when
computing the confidence interval for the solution, they can be
neglected when they are small in comparison with the perturbations
of the forcing function.

The comparison of speed is somewhat subjective, since we have no
insights into how much function-call-overhead is involved in the
MATLAB implementation; nevertheless, it does show the potential
speed of the new approach. This test demonstrates the ability of
the new method to compute solutions to the ODE at arbitrary given
nodes with a very high accuracy.
\subsection{Test B: Alternative Node Placement for Initial Value Problem 1}
\begin{figure}
  \begin{minipage}[t]{0.49\columnwidth}
    \centering
    \includegraphics[width=65mm]{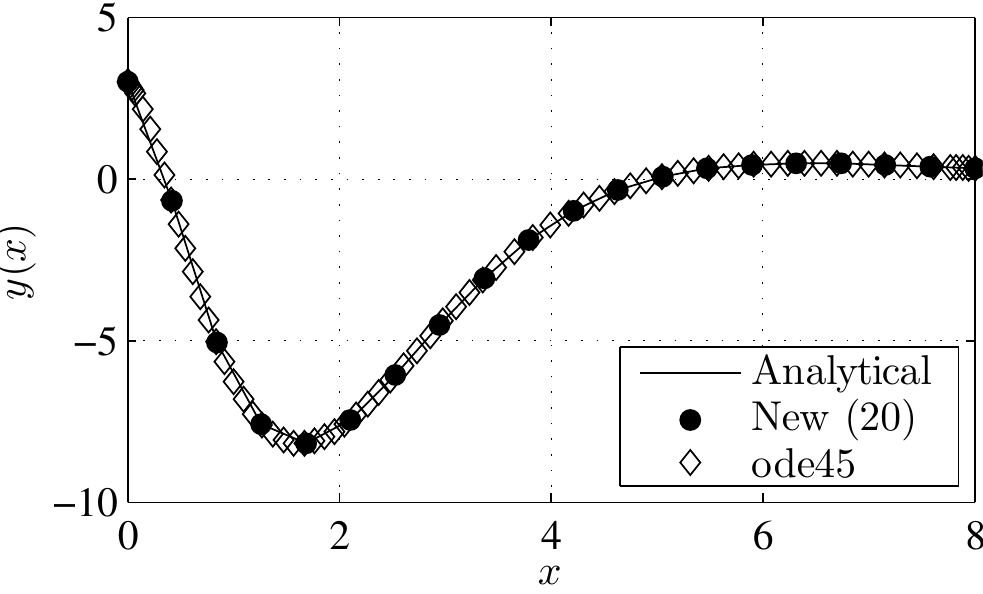}
    \caption{\revtwo{The plot shows the analytic solution from Eqn.~(\ref{eqn:testAanalytical}), the solution with the new method with $l_s=9$ and $n=20$ as well as the solution by a Runge-Kutta {\tt ode45} method with $n=77$.}}
    \label{fig:ivpExp3c}
  \end{minipage}
  \hfill
  \begin{minipage}[t]{0.49\columnwidth}
    \centering
    \includegraphics[width=65mm]{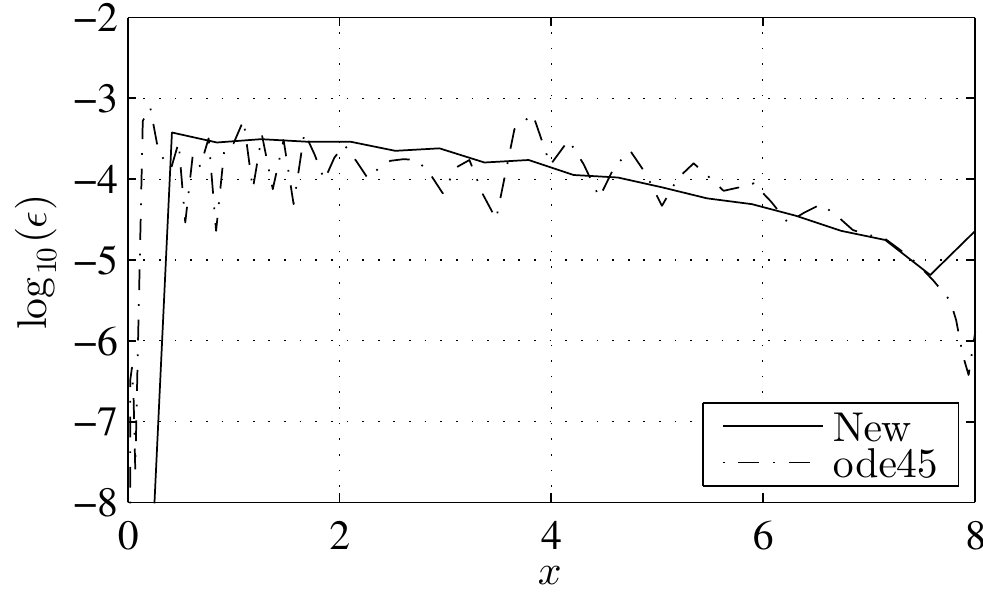}
\caption{\revtwo{Residual errors: difference of the new numerical solution with $n=20$ vs. the analytical solution and the solution by a Runge-Kutta {\tt ode45} method with $n=77$ vs. the analytical solution. }}
    \label{fig:ivpExp3d}
  \end{minipage}
\end{figure}

In this test the same ODE is solved as in Test A, however, a reduced number of $n = 20$ evenly spaced nodes has been selected for the new method since systems sampling in time in general use even spacing. The results are shown in Fig.~\ref{fig:ivpExp3c} and \ref{fig:ivpExp3d}. \revtwo{The new method achieves the same solution quality, in terms of accuracy, as the {\tt ode45} method, however with a significantly reduced number of nodes.} This corresponds to an accurate solution of the inverse problem with a small number of sensors.
\subsection{Test C: Initial Value Problem 2} \label{secIVP2}
\begin{figure}
  \begin{minipage}[t]{0.49\columnwidth}
    \centering
    \includegraphics[width=65mm]{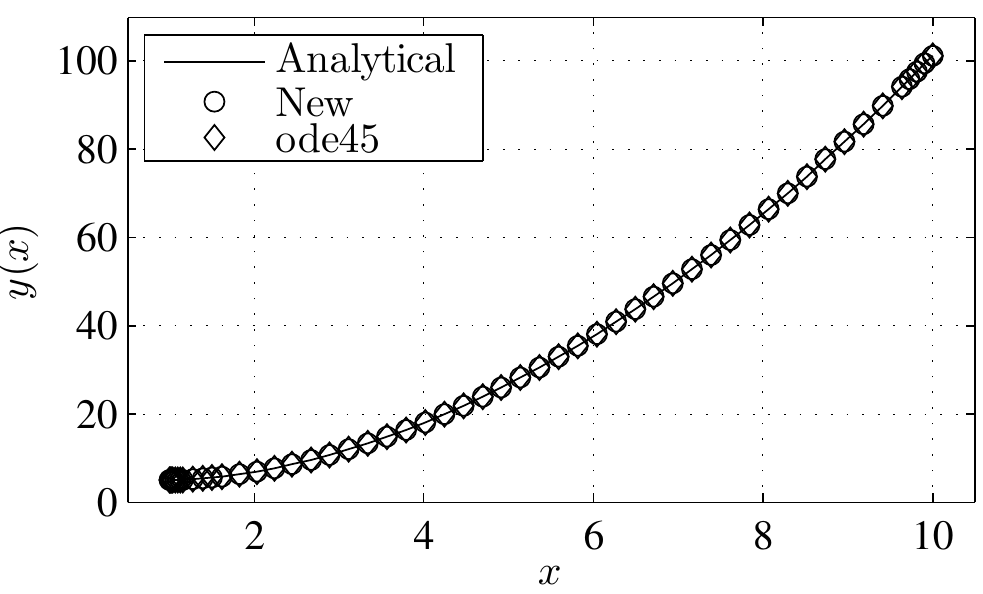}
    \caption{\revtwo{The plot shows the analytic solution from Eqn.~(\ref{eqn:testBanalytical}), the solution with the new method with $l_s=15$ and $n = 69$ evenly placed nodes as well as the solution by a Runge-Kutta {\tt ode45} method with $n = 69$ variably placed nodes.}}
    \label{fig:ivpExp2a}
  \end{minipage}
  \hfill
  \begin{minipage}[t]{0.49\columnwidth}
    \centering
    \includegraphics[width=65mm]{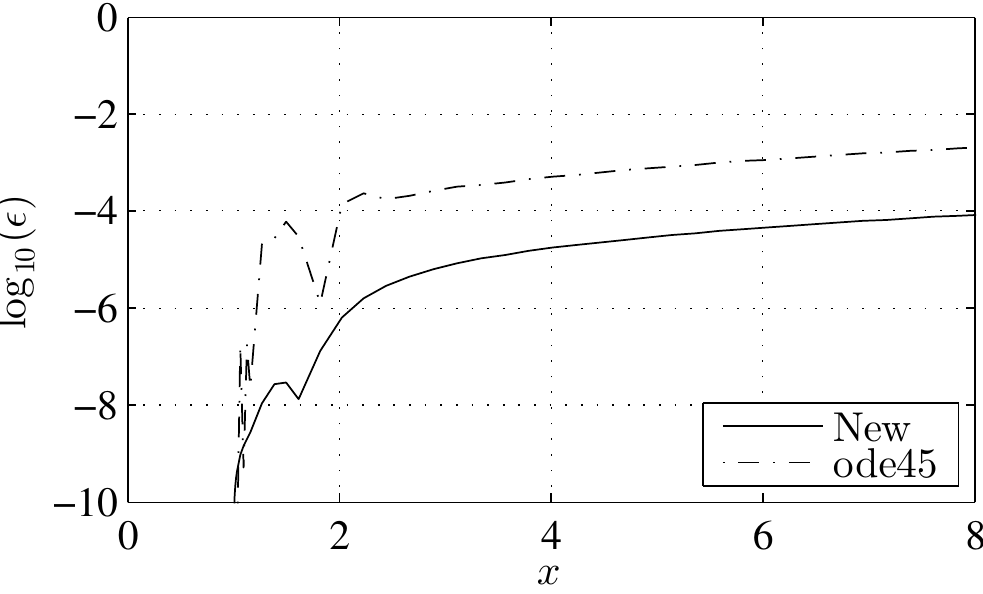}
    \caption{\revtwo{Residual errors: difference of the new numerical solution vs. the analytical solution and the solution by a Runge-Kutta {\tt ode45} method vs. the analytical solution.}}
    \label{fig:ivpExp2b}
  \end{minipage}
\end{figure}

The second example is a second order ($m=2$) ODE with variable
coefficients $a_i(x)$ and $p=2$ constraints. This demonstrates the ability of the method to deal
with variable coefficients and with solutions which are irrational
functions. The equation is,
\begin{align}\label{eqn:IVP2}
    & 2\,x^2\,y'' - x\,y' - 2 y = 0 \eqtext{given} \\
    & y(1) = 5,
    \hspace{5mm}
    y'(1) = 0 \notag
\end{align}
in the interval $1 \leq x \leq 10$. The analytical solution to
this equations is
\begin{equation} \label{eqn:testBanalytical}
    y(x) = x^2 + \frac{4}{\sqrt{x}}.
\end{equation}
The solution's appearance would not suggest that this is a
demanding problem. However, the analytical solution is the sum of
a polynomial and an irrational function. Computing good estimates
for the derivatives of such functions can require a high degree of
polynomial approximation. The solution obtained using the new
method, the analytical solution and the result of the {\tt ode45}
solver are shown in Fig.~\ref{fig:ivpExp2a}. The high density of
nodes at the start of the interval produced by the {\tt ode45} method
indicates that the method \revfive{required disproportionately many steps for} finding a
solution with sufficient accuracy. The new method is once again
more accurate than the {\tt ode45} solver.
\subsection{Test D: Selecting a Support Length for Initial Value Problem 2}
\begin{figure}[ht]
    \centering
    \includegraphics[width=70mm]{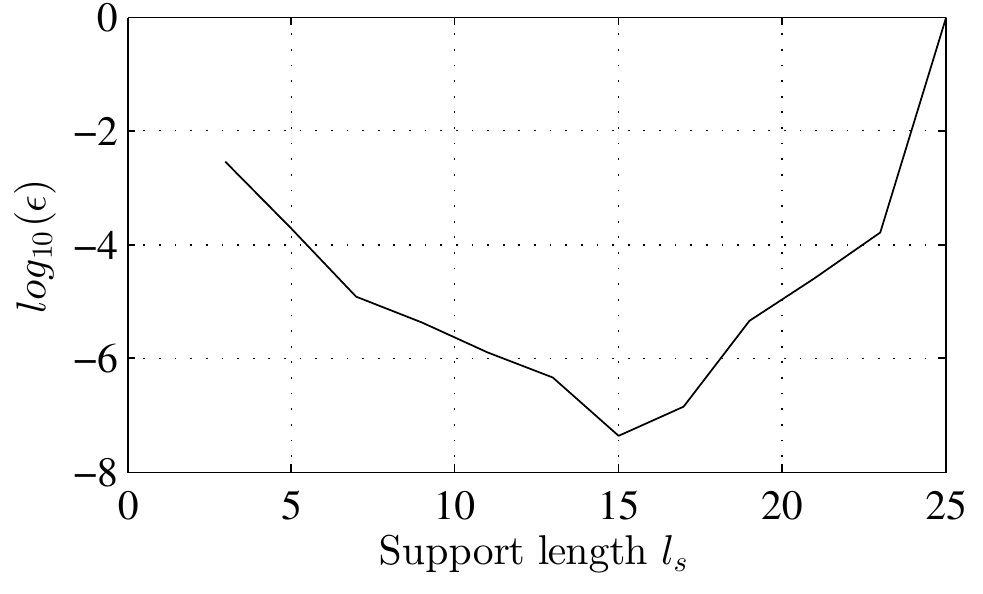}
    \caption{The relative error $\epsilon$ has a minimum for $l_s = 15$.}
    \label{fig:ivpExp2c}
\end{figure}

In this test, the same ODE is solved as in Test C. As pointed out
in Section~\ref{sec:previous}, there is an issue in selecting the
support length $l_s$ (or degree) of the local approximation for
the differentiating matrix $\M{D}$. This matrix and the linear
differential operator $\M{L}$ have been implemented for all odd
support lengths in the range $3 \leq l_s \leq 25$ and the IVP was
solved for each of these implementations. The relative error was
computed for the corresponding solutions as
\begin{equation}
    \epsilon(l_s) = \frac{{|\V{y}_a - \V{y}|}_2}{{|\V{y}_a|}_2},
\end{equation}
where $\V{y}_a$ is the sampled analytical solution and $\V{y}$ is
the solution computed with the new method. The
$\log_{10}(\epsilon)$ vs. $l_s$ is shown in
Fig.~\ref{fig:ivpExp2c}. This result shows that there is a minimum
in the relative error for $l_s = 15$, indicating that there is a justification for implementing
local approximation to derivatives for \revfive{specific problems with high numbers of nodes}\footnote{Many
books~\cite{burden2005,lapidus1999,strikwerda200} discuss the
possibility of implementing approximations of higher degree;
however, they never actually show comparative \revfive{numerical results for practical problems}.}. The
dependence of $\epsilon$ on $l_s$ is a function of the equation
being solved. There will be no solution that is optimal for all
cases. With the proposed method the necessary $l_s$ is determined
during the preparatory computations and not at run-time.
\subsection{Test E: Inverse 3-Point Boundary Value Problem} \label{secBVP}
The following test is an over constrained first order ($m=1$)
$3$-point inverse BVP\footnote{\revfive{At least one of the
constraints is interior, since the constraints are not restricted
to the boundaries.}}; however, with $p=4$ constraints at $3$
locations on the abscissae. It belongs to the class of inverse
multi-point BVPs\footnote{Although we have been able to find a
number of publications on methods relating to multi-point
BVPs~\cite{Welsh1980,Agarwal2003}, there is very little literature
available on inverse multi-point BVPs, e.g.~\cite{Kurylev1993}.
There are no general approaches available at the present time.}.
The constraints implemented are both: homogeneous and
non-homogeneous; as well as Dirichlet and Neumann boundary
conditions. This example has been chosen to demonstrate the
numerical efficiency and behavior of the method with respect to a
perturbed inverse BVP. Furthermore, it demonstrates the ability of
the algebraic framework to  deal with generalized constraints.
Synthetic data is produced for a function and its analytic
derivatives, in this manner the result of the reconstruction can
be compared with the function from which the data was derived.

\begin{figure}[ht]
    \centering
    \includegraphics[width=70mm]{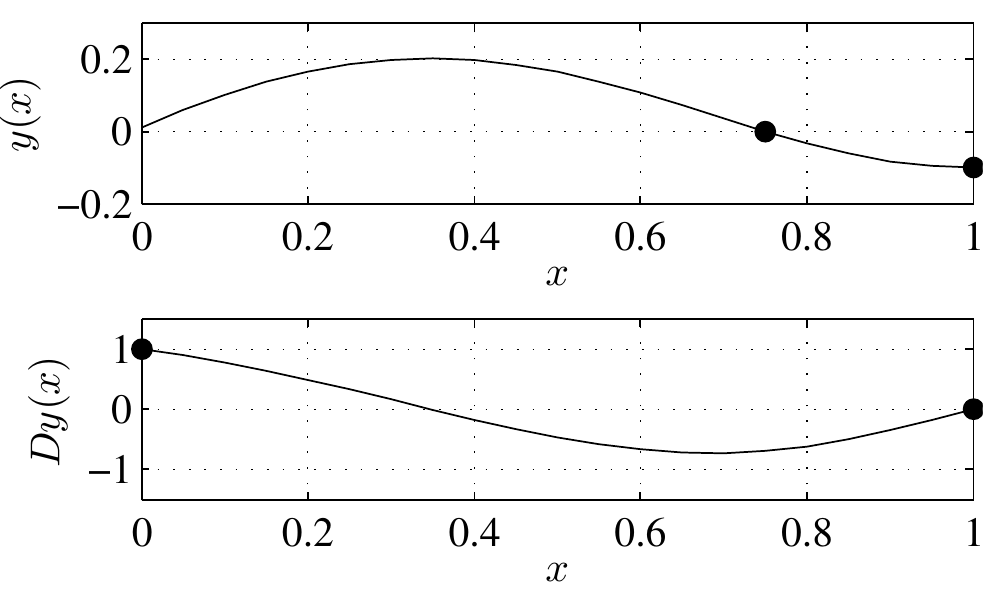}
    \caption{\revtwo{First order ($m=1$) constrained test function and its analytical
    derivatives: this function
    fulfills all $p=4$ constraints specified in
    Eqn.~(\ref{eqn:BVP1c1}) and (\ref{eqn:BVP1c2}).
    The homogeneous and non-homogeneous Dirichlet and Neumann
    constraints are marked at the three locations $[0,0.7895,1]$ in the top and bottom plots respectively.}}
    \label{fig:BVP1a}
\end{figure}
The problem being considered is to reconstruct a curve $\V{y}$ from
multiple local measurements of the curve's gradients $\V{g}$ while
fulfilling a set of constraints $\MT{C} \V{y} = \V{d}$, which are not restricted to
the ends of the support, i.e., inner constraints are also
present. Furthermore, the measurements are perturbed by white noise. The ODE is
\begin{equation}
    y'(x) = g(x) + \epsilon,
\end{equation}
where $\epsilon$ is the error caused by the forcing function's
perturbation. This equation is subject to the homogeneous and non-homogeneous Dirichlet boundary
conditions,
\begin{align}
    y(0.7895) = 0 \eqtext{and} y(1) = -0.1 \label{eqn:BVP1c1}
\end{align}
as well as the non-homogeneous and homogeneous Neumann boundary conditions,
\begin{align}
    y'(0) = 1 \eqtext{and} y'(1) = 0.\label{eqn:BVP1c2}
\end{align}

A synthetic test function which fulfills these conditions was
generated by combining an arbitrary polynomial of $4^{\text{th}}$
degree,
\begin{equation}
    y(x) = 1.1 x^{4} + 0.4 x^{3} + 0.5 x^{2} - 1.2 x - 0.3,
\end{equation}
with the four constraints, this yields an $8^{\text{th}}$ degree
polynomial\footnote{The theory behind this computation can be
found in \cite{harker2013poly}.} which also fulfills the
constraints:
\begin{align}
    y_c(x) =
 &- 0.46985 x^{8} + 0.41127 x^{7} + 0.34891 x^{6} + 0.03827 x^{5} \notag\\
 &+ 1.0323 x^{4} - 1.5886 x^{3} - 0.88426 x^{2} + x +
 0.011895\label{eqn:constPoly}.
\end{align}
The first derivative of Eqn.~(\ref{eqn:constPoly}) can be computed
analytically, making it a suitable test function for constrained
curve reconstruction from gradients. The function $g(x)$, its
analytical gradient $g'(x)$ and the constraints are shown in
Fig.~\ref{fig:BVP1a}.

\begin{figure}
  \begin{minipage}[t]{0.49\columnwidth}
    \centering
    \includegraphics[width=65mm]{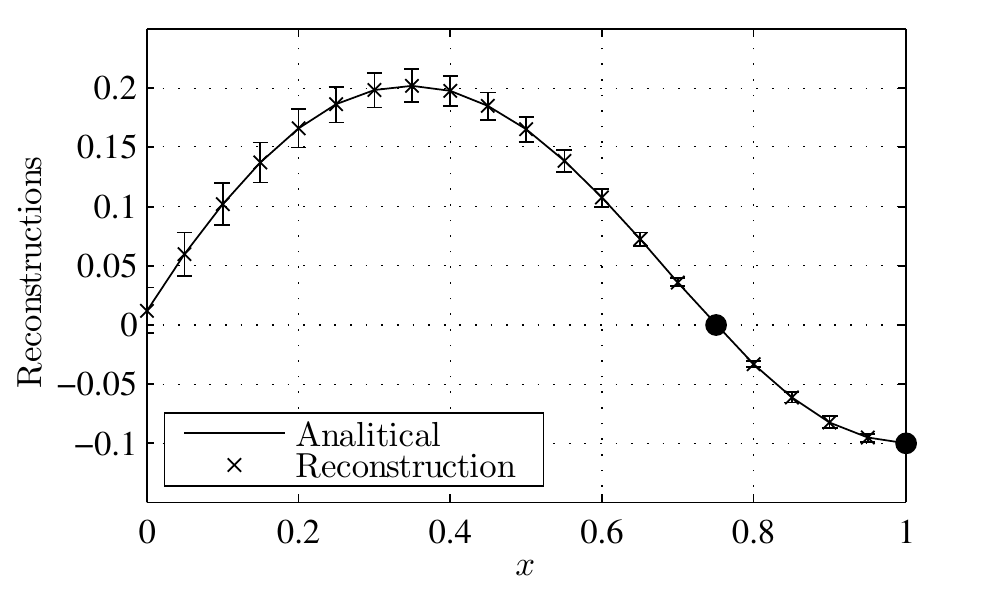}
    \caption{Comparison of the analytical solution and the
    result of the reconstruction from the perturbed gradients.
    The two Dirichlet boundary conditions are marked on the
    reconstruction, the two Neumann conditions are not shown.
    The error bars correspond to the standard deviations, i.e.,
    an estimate for the \revtwo{$68.3 \%$ confidence interval}.
    They are obtained from a Monte Carlo simulation
    with $k = 10000$ iterations. The error bars have been
    magnified by a factor of $10$ to increase the visibility.}
    \label{fig:BVP1b}
  \end{minipage}
  \hfill
  \begin{minipage}[t]{0.49\columnwidth}
    \centering
    \includegraphics[width=65mm]{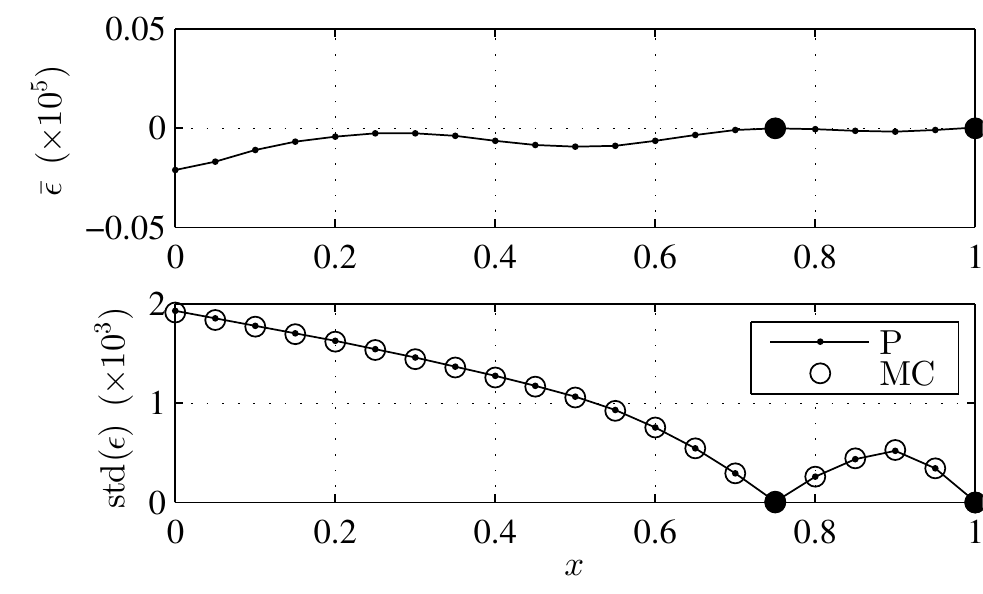}
    \caption{Top: Bias of the reconstruction, i.e., the difference
    between the mean reconstruction from the $k = 10000$ Monte
    Carlo iterations and the analytical solution (the result is
    scaled by $10^5$ to make the error visible.).
    Bottom: The standard deviation of the
    reconstruction as predicted (P) by the covariance propagation
    according to Eqn.~(\ref{eqn:statsFinal}) and the results from
    the Monte Carlo simulations (MC) (the result is scaled by $10^3$).}
        \label{fig:BVP1c}
  \end{minipage}
\end{figure}
The analytical gradient is evaluated at $n = 21$
points\footnote{For example, these would correspond to the
positions of the inclinometers on a structure being monitored.}.
Then in a Monte Carlo simulation, with $k = 10000$ iterations,
the gradients are perturbed by artificial Gaussian noise with a
standard deviation of $1 \%$ of the maximum value of $D y(x)$. For
each simulation, a reconstruction is performed with the appropriate
$\M{M}$ and $\V{y}_h$ and the statistics are
computed. The result of solving this problem using the proposed
method is shown in Fig.~\ref{fig:BVP1b}, together with the error
bars corresponding to the standard deviation of the reconstructed
values observed in the Monte Carlo simulation. The error bars have
been magnified by a factor of $10$ to increase the visibility.

In Fig.~\ref{fig:BVP1c}, the bias of the reconstruction is shown,
i.e., the difference between the analytical solution and the mean
of the Monte Carlo simulations. Additionally, the standard
deviation of the result as predicted by
Eqn.~(\ref{eqn:statsFinal}) and the results of the Monte Carlo
simulation are compared. These results verify that the solution
is, for all intents and purposes, bias free and the predicted
uncertainty is correct. The method has been successfully applied
to an inverse three-point boundary value problem, with two
Dirichlet and two Neumann boundary conditions. Not only is the
problem solved, but in addition the uncertainty of the solution is
delivered by the new method.
\subsection{Summary of the Numerical Testing}
The implications of the above test can be summarized as follows:
\begin{enumerate}
    \item The new method is capable of solving both perturbed and
    unperturbed inverse problems, including initial-,
    inner- and boundary value problems.
    \item The method enables the formulation and solution of
    problems with constraints on arbitrary derivatives of the
    solution function.
    Consequently, both Dirichlet and Neumann boundary conditions
    can be dealt with. The constraints can be both homogeneous ($d=0$) and
    non-homogeneous ($d \neq 0$).
    \item The method exhibits significantly smaller numerical
    errors than the Runge-Kutta {\tt ode45} approach, while
    being significantly faster. Reducing the error bound for the
    {\tt ode45} will improve the numerical accuracy,
    but at the expense of computational effort.
    The numerical errors are so small
    that they can be neglected when solving inverse problems where the
    perturbation of the forcing function is significant.
    \item The separation of the solution into a preparatory (offline) and
    run-time (online) computation makes the method suitable for embedding in
     real-time systems.
\end{enumerate}
\section{Automatic Code Generation}
\revfour{The aim of the code generation is, given the definition of specific inverse problem in terms of a measurement model (see Section~\ref{sec:measurementModel}), to automatically generate the code required to solve the problem on an embedded computing system. That is, the problem is defined as a symbolic definition of a differential equation together with a suitable set of constraints and a source of data. The system will then solve any ODE, regardless of its nature (IVP, BVP, etc.) from this specification. A similar concept of automatic code generation (ACG) for the embedding of convex optimization was explained in literature~\cite{Mattingley2012}. Furthermore, it was shown, that numerical ODE solvers can be deployed on FPGAs using VHDL in \cite{Huang2013}.}
During MBD, the system is designed on an abstract model level based on the system's requirements
while ensuring the consistency of the system's physical
representation. ACG is the task of converting the models and their
algorithms to usable code, effectively automating the
time-consuming and error-prone process of low-level programming.
Basically, there are three low-level target languages for embedded
systems: multi-purpose \textit{ANSI-C code}, which is the focus of
this article; \textit{hardware description language (HDL)} for field
programmable gate arrays (FPGA) or application specific integrated
circuits (ASIC); and IEC 61131-3 compliant languages such as
\textit{structured text (ST)} for programmable logic controllers
(PLC).

Code deployment is the integration of code on the embedded
systems. In most cases, the architecture of the development (host)
system (x86 or x64) is largely different from the embedded
(targeted) system (ARM, ATmel, etc.). There are two approaches for
solving this issue \textit{compilation of code on the target
system} if an OS and an appropriate compiler is present; or
\textit{cross compilation} on the host system via processor
virtualization, a popular tool to perform this task is the LLVM
compiler infrastructure.

After MBD is complete, so called production code is generated. The
process strips out all parameters needed during testing and
optimizes the code for performance (low memory usage, high
computational speed) or safety (data consistency, robust
algorithms).
\subsection{Embedded Target Hardware}
%
%
The goal is to show that even highly abstract and complex mathematical models are deployable on the simplest embedded hardware, demonstrating the scalability of the method. The open-source \textit{Raspberry Pi} or the proprietary \textit{BeagleBone Black} are popular entry-level embedded systems for target programming. The \textit{WAGO PFC-200} is an IEC 61131-3 compliant industrial PLC with open source software.  These three systems are based on $32$-bit ARM processors and they run an embedded Linux derivative as operating system (OS); in the case of the WAGO device it's a real-time OS. The automated resource management is the main advantage of embedded systems with an OS. The low end of the systems is represented by the fully open source \textit{Arduino Uno} platform. The utilized Atmel $8$-bit AVR RISC-based ATmega328 microcontroller has no dedicated OS, the program logic is directly stored on the chip's $32 \unit{kB}$ flash memory as firmware. \revfour{The Arduino Uno has been chosen for experimental PIL testing, see Section~\ref{sec:exptest}. A laboratory experiment featuring the BeagleBone Black is presented in Section~\ref{sec:laborTest}.}
%
%
%
%
\subsection{MBD Software for Code Generation}
Most engineering and scientific software for designing mathematical models has the functionality to automatically generate standard ANSI-C code from its application-specific syntax, e.g., LabVIEW, Maple or Mathematica. This is usually necessary, because most industrial controllers are only programmable with C.
%
%
%
A short survey on tools for ACG has been given by Rafique et al.~\cite{rafique2013}. This article is focused on the usage of MATLAB and its Coder toolbox, because it is the standard software for mathematical MBD. Code generation fully supports linear algebra. \revfour{Nevertheless, the presented approach is so simple, that a C code parser could be implemented manually without much effort.}
Two test cases confirmed the correct functionality of algebraic functions generated by MATLAB Coder: C code was generated for QR decomposition and singular value decomposition. Compiling the code with Microsoft Visual Studio 2010 for the $32$-bit host development system delivered an executable program, which successfully validated the code via SIL verification. Deploying the same code on the Arduino Uno confirmed the correct functionality via PIL verification.
\begin{figure}[ht] 
  \centering
  \includegraphics[width=140mm]{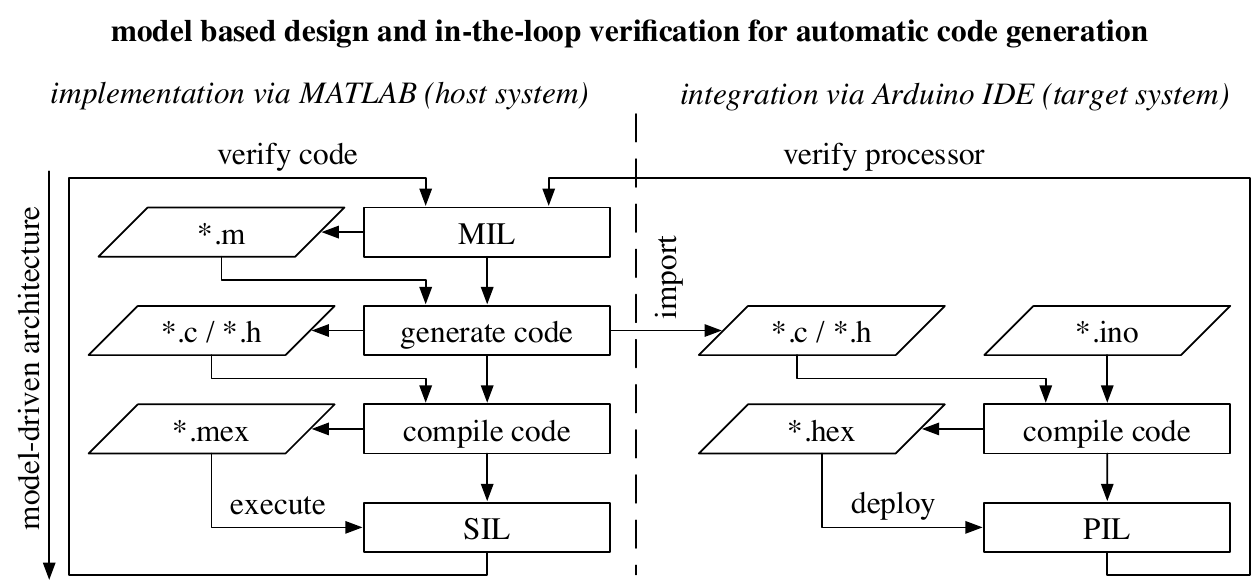}\\
  \hspace{2mm}\\
  \caption[Model Based Design Process]{MBD is an iterative approach. Each step requires verification to ensure that the (sub)system's requirements are met. In this article, model-in-the-loop (MIL) verifies the correctness of the algebraic framework on model level, software-in-the-loop (SIL) verifies the functional equivalence of the generated C code on code level and processor-in-the-loop (PIL) verifies the correct computation on the employed microcontroller on binary level. This graphic shows the process for the Arduino platform.}
  \label{fig:MBD}
\end{figure}
\subsection{Targeting and Verification Process}
The ACG process is completely general, the illustration in Fig.~\ref{fig:MBD} shows the procedure for the Arduino platform. Following steps must be carried out:

\textit{Model-in-the-loop (MIL):} The system is identified, designed and simulated on abstract model level in an artificial environment, producing MATLAB model code (*.m). This is an efficient way to estimate model parameters with varying configurations. This includes the determination of the optimal support length $l_s$, computation of the constrained linear differential operator $\M{M}$ and the homogeneous solution $\V{y}_h$. Furthermore, changes in the requirements can easily be implemented in this early design stage. 

\textit{Code generation:} The MATLAB Coder toolbox is a sophisticated parser engine. It converts the model code (*.m) into C code (*.c) and the associated header files (*.h). 

\textit{Code compilation for SIL:} MATLAB Coder features the ability to replace model function calls with calls for MATLAB executables (*.mex). Such functions are wrappers around compiled C code, which can be directly called from the MATLAB development environment.

\textit{Software-in-the-loop (SIL):} The model and the generated C code must be functionally equivalent, i.e., a certain input must deliver the same output on all abstraction layers. This is especially relevant when the target language misses certain features of the model language, e.g. shorter bit-lenghts of variable types or no support for floating point operations. The consistency of the model must be ensured on all levels.

\textit{Code compilation for PIL:} The C code (*.c/*.h) is imported into the Arduino IDE. The header (*.h) files must be included in the Arduino project's main (*.ino) file. The C code is cross-compiled for the Arduino platform delivering a (*.hex) file, which is stored directly on the ATmega328's flash memory as firmware.

\textit{Processor-in-the-loop (PIL):} The code runs on the embedded real-time system. The outcome is not necessarily the same as during simulation, because the hardware platform used during MIL and SIL is different from the PIL target.
\section{Software- and Processor-in-the-Loop Testing} \label{sec:exptest}
In Section~\ref{secTesting}, the viability of the new method was shown during MIL. In this section, the test cases are directly executed on the Arduino Uno for PIL verification. The microcontroller features $2 \unit{kB}$ SRAM and a processing power of $16$ million instructions per second (MIPS). The $23$ general purpose I/O lines, the $6$-channel $10$-bit A/D converter and the operating voltage of $1.8-5.5 \unit{V}$ makes it a well suited setup for acquiring and processing sensor data. 
The problem size must be scaled down in order to fit the Arduino Uno's limited system resources. An Arduino Uno double variable requires $4 \unit{B}$ of memory, so theoretically, the ATmega328 chip stores up to $512$ double variables in its memory of $2 \unit{kB}$. Obviously, operations and other variables also require memory space, therefore the problem size has been shrunk to $10$ input signals. This corresponds in means of problem size to the Arduino Uno's $6$ analog I/O ports, which are usable to connect sensors to the device. However, the problem classes are still the same. 

The constrained linear differential operator $\M{M}$ has then a size of $ (10 \times 10)$ and the homogeneous solution vector $\V{y}_h$ has a size of $ (10 \times 1)$. Both, $\M{M}$ and $\V{y}_h$, are computed a-priori during offline calibration. Only the measurement vector $\V{g}$ with size $(10 \times 1)$ changes its values from one measurement to the next. Consequently, the result of the online computation, i.e. the solution vector $\V{y}$, is of size $(10 \times 1)$.
\subsection{Initial Value Problem 1}
\begin{figure}[h]
  \begin{minipage}[t]{0.49\columnwidth}
    \centering
    \includegraphics[width=65mm]{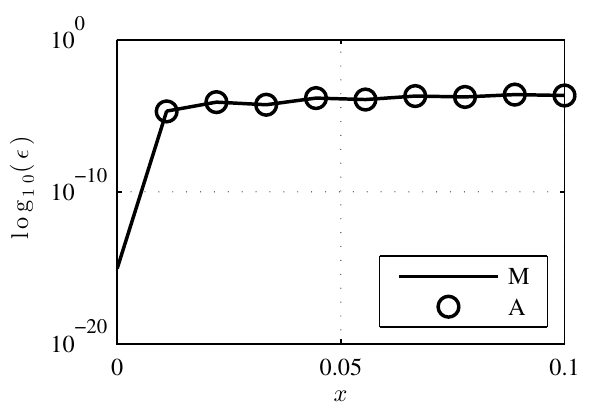}
    \caption{\revtwo{The plot shows the error between the analytic solution and MATLAB's solution (M), the error norm is $|\V{\epsilon}|_2 = 4.6771 \, 10^{-4}$, as well as the error between the analytic solution and Arduino's solution (A), the error norm is $|\V{\epsilon}|_2 = 4.6749 \, 10^{-4}$.}}
    \label{fig:arduinoIVP3log}
  \end{minipage}
  \hfill
  \begin{minipage}[t]{0.49\columnwidth}
    \centering
    \includegraphics[width=65mm]{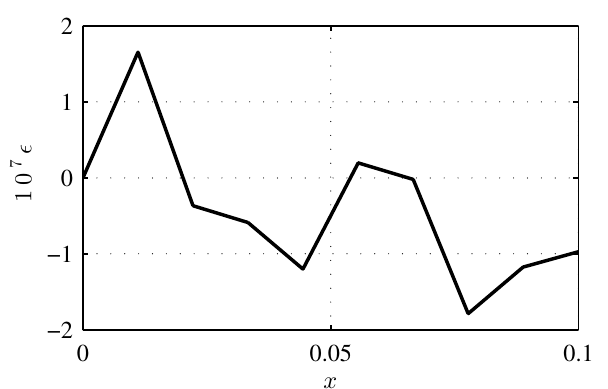}
    \caption{\revtwo{PIL verification: the difference between MATLAB's solution and Arduino's solution is shown, the error norm is $|\V{\epsilon}|_2 = 3.1999 \, 10^{-7}$. The result is scaled by $10^7$ to increase the visibility.}}
    \label{fig:arduinoIVP3}
  \end{minipage}
\end{figure}

The test case in Section \ref{secIVP1} has been modified to have $n = 10$ evenly spaced nodes in the interval $ 0 \leq x \leq 0.1$ with a support length of $l_s = 5$. The computation time on the Arduino Uno is $t = 1.788 \unit{ms}$, i.e., a sample rate of $>500 \unit{Hz}$ is possible. The error plots of the numerical computations are shown in Fig.~\ref{fig:arduinoIVP3log} and \ref{fig:arduinoIVP3}.
\subsection{Initial Value Problem 2}
\begin{figure}[h]
  \begin{minipage}[t]{0.49\columnwidth}
    \centering
    \includegraphics[width=65mm]{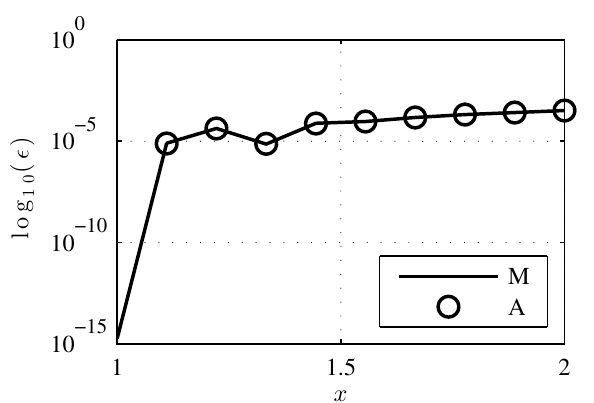}
    \caption{\revtwo{The plot shows the error between the analytic solution and MATLAB's solution (M), the error norm is $|\V{\epsilon}|_2 = 4.9541 \, 10^{-4}$, as well as the error between the analytic solution and Arduino's solution (A), the error norm is $|\V{\epsilon}|_2 = 4.9568 \, 10^{-4}$.}}
    \label{fig:arduinoIVP2log}
  \end{minipage}
  \hfill
  \begin{minipage}[t]{0.49\columnwidth}
    \centering
    \includegraphics[width=65mm]{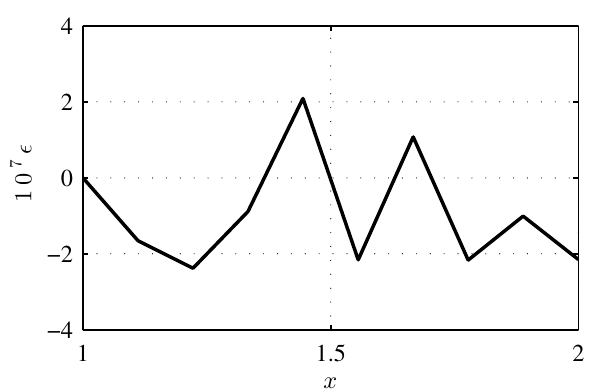}
    \caption{\revtwo{PIL verification: the difference between MATLAB's solution and Arduino's solution is shown, the error norm is $|\V{\epsilon}|_2 = 5.4520 \, 10^{-7}$. The result is scaled by $10^7$ to increase the visibility.}}
    \label{fig:arduinoIVP2}
  \end{minipage}
\end{figure}

The test case in Section \ref{secIVP2} has been modified to have $n = 10$ evenly spaced nodes in the interval $ 1 \leq x \leq 2$ with a support length of $l_s = 5$. The computation time on the Arduino Uno is $t = 1.228 \unit{ms}$, i.e., a sample rate of $>800 \unit{Hz}$ is possible. The error plots of the numerical computations are shown in Fig.~\ref{fig:arduinoIVP2log} and \ref{fig:arduinoIVP2}
\begin{figure}
  \begin{minipage}[t]{0.49\columnwidth}
    \centering
    \includegraphics[width=65mm]{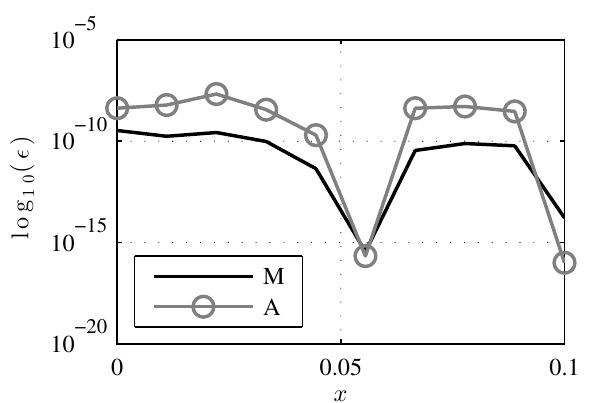}
    \caption{\revtwo{The plot shows the error between the analytic solution and MATLAB's solution (M), the error norm is $|\V{\epsilon}|_2 = 4.7958 \, 10^{-10}$, as well as the error between the analytic solution and Arduino's solution (A), the error norm is $|\V{\epsilon}|_2 = 2.3902 \, 10^{-8}$.}}
    \label{fig:arduinoBVPlog}
  \end{minipage}
  \hfill
  \begin{minipage}[t]{0.49\columnwidth}
    \centering
    \includegraphics[width=65mm]{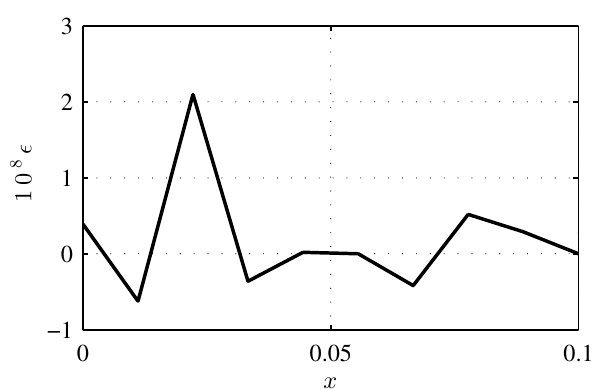}
    \caption{\revtwo{PIL verification: the difference between MATLAB's solution and Arduino's solution is shown, the error norm is $|\V{\epsilon}|_2 = 2.3673 \, 10^{-8}$. The result is scaled by $10^8$ to increase the visibility.}}
    \label{fig:arduinoBVP}
  \end{minipage}
\end{figure}
\subsection{Inverse 3-Point Boundary Value Problem}
The test case in Section \ref{secBVP} has been modified to have $n = 10$ evenly spaced nodes in the interval $ 0 \leq x \leq 0.1$ with a support length of $l_s = 5$. New constraints have been defined to conserve the test case's characteristics:
\begin{align}
    y(0.0556) & =  0, & y(0.1)    & = -0.1,\\
    D\,y(0)   & =  1, & D\,y(0.1) & =  0.
\end{align}
The computation time on the Arduino Uno is $1.796 \unit{ms}$, i.e., a sample rate of $>550 \unit{Hz}$ is possible. The error plots of the numerical computations are shown in Fig.~\ref{fig:arduinoBVPlog} and \ref{fig:arduinoBVP}. In contrast to the previous test cases, here the SIL verification delivered an error vector with norm of $|\V{\epsilon}|_2 = 3.3866 \, 10^{-14}$, i.e., the C code's result is slightly different than the MATLAB model code's result.
\section{Laboratory Testing} \label{sec:laborTest}
\begin{figure}[h]
  \centering
  \includegraphics[width=140mm]{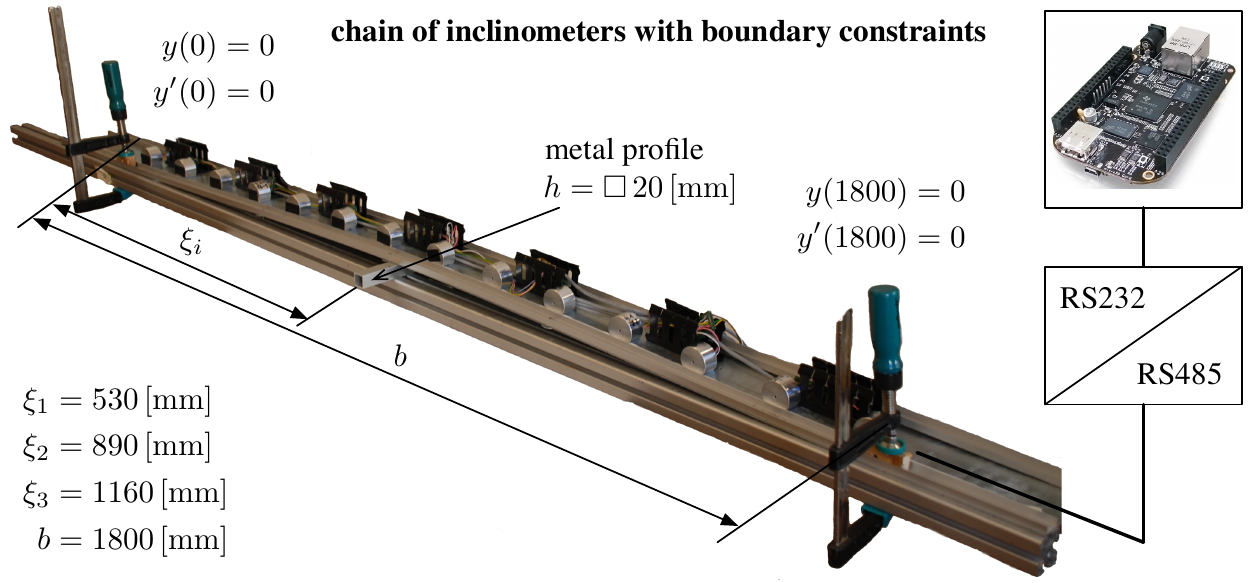}\\
  \hspace{2mm}\\
  \caption[Inclinometer System]{\revfour{The illustration shows the chain of inclinometers mounted on a flexible structure. Each of the $14$ sensors is connected to an industrial RS-485 bus. This bus is converted to a RS-232 serial interface, which enables the connection of the BeagleBone Black. The $2$ screw clamps force the homogeneous boundary values at the structure's ends.}}
  \label{fig:inclinometerSystem}
\end{figure}
\begin{figure}
  \begin{minipage}[t]{0.49\columnwidth}
    \centering
    \includegraphics[width=65mm]{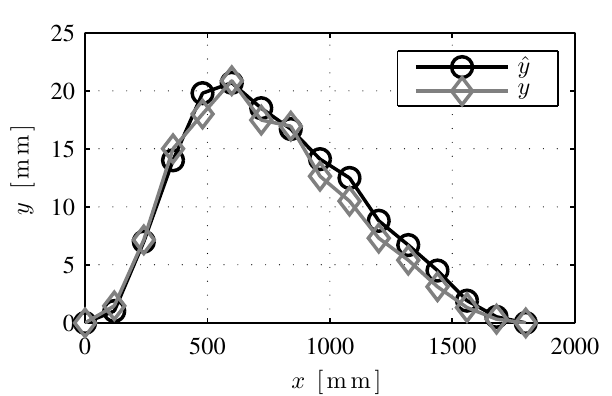}
    \caption{\revfour{The reference values $\hat{\V{y}}$ and the computed curve $\V{y}$ for the metal profile placed at $\xi_1 = 530 \unit{mm}$ are shown.}}
    \label{fig:inclinometerCompare1}
  \end{minipage}
  \hfill
  \begin{minipage}[t]{0.49\columnwidth}
    \centering
    \includegraphics[width=65mm]{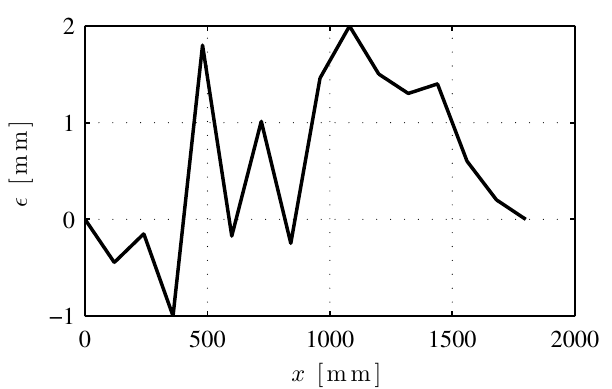}
    \caption{\revfour{The difference between both values $\V{\epsilon} = \hat{\V{y}} - \V{y}$ is shown, the error norm is $|\V{\epsilon}|_2 = 4.245 \unit{mm}$.}}
    \label{fig:inclinometerError1}
  \end{minipage}
\end{figure}
\begin{figure}
  \begin{minipage}[t]{0.49\columnwidth}
    \centering
    \includegraphics[width=65mm]{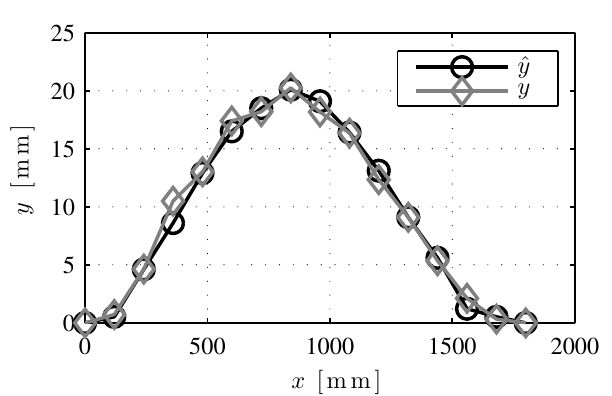}
    \caption{\revfour{The reference values $\hat{\V{y}}$ and the computed curve $\V{y}$ for the metal profile placed at $\xi_2 = 890 \unit{mm}$ are shown.}}
    \label{fig:inclinometerCompare2}
  \end{minipage}
  \hfill
  \begin{minipage}[t]{0.49\columnwidth}
    \centering
    \includegraphics[width=65mm]{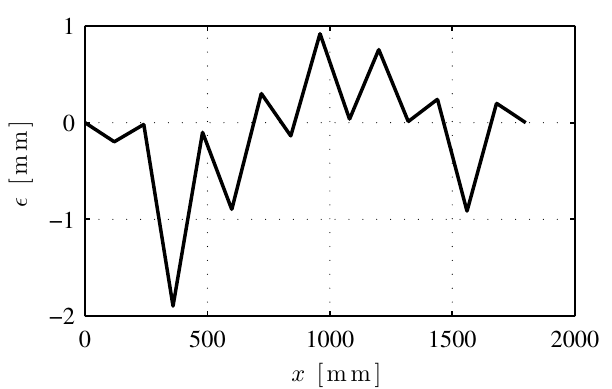}
    \caption{\revfour{The difference between both values $\V{\epsilon} = \hat{\V{y}} - \V{y}$ is shown, the error norm is $|\V{\epsilon}|_2 = 2.636 \unit{mm}$.}}
    \label{fig:inclinometerError2}
  \end{minipage}
\end{figure}
\begin{figure}
  \begin{minipage}[t]{0.49\columnwidth}
    \centering
    \includegraphics[width=65mm]{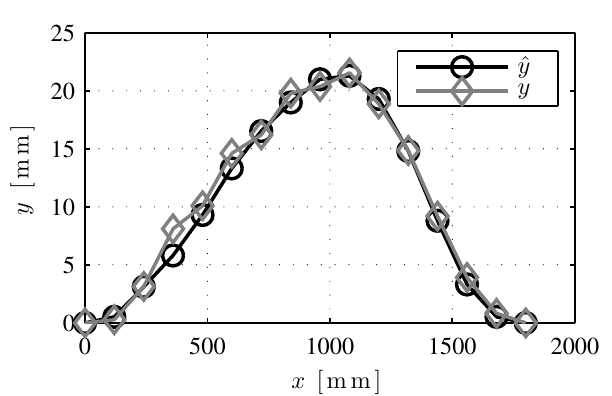}
    \caption{\revfour{The reference values $\hat{\V{y}}$ and the computed curve $\V{y}$ for the metal profile placed at $\xi_3 = 1160 \unit{mm}$ are shown.}}
    \label{fig:inclinometerCompare3}
  \end{minipage}
  \hfill
  \begin{minipage}[t]{0.49\columnwidth}
    \centering
    \includegraphics[width=65mm]{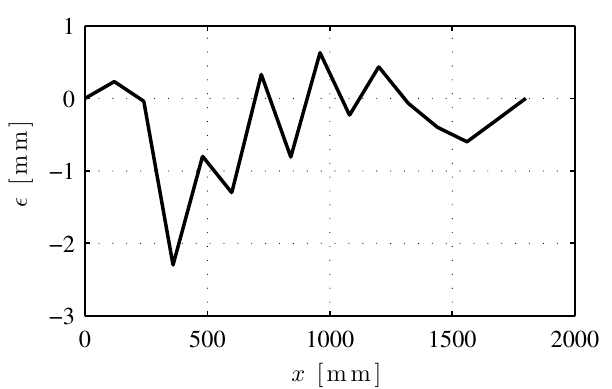}
    \caption{\revfour{The difference between both values $\V{\epsilon} = \hat{\V{y}} - \V{y}$ is shown, the error norm is $|\V{\epsilon}|_2 = 3.114 \unit{mm}$.}}
    \label{fig:inclinometerError3}
  \end{minipage}
\end{figure}
\revfour{
The introduced algebraic model is tested on a laboratory setup, see Fig.~\ref{fig:inclinometerSystem}. A chain of equally spaced one-dimensional inclinometers is mounted on a $b=1.8 \unit{m}$ long flexible structure. The arrangement consists of $14$ sensors with an additional $2$ screw clamps, effectively forcing $2$ pairs, i.e. $p=4$, of homogeneous Dirichlet and Neumann constraints at the structure's ends. These leads to a total of $n=16$ points for the computation, i.e., the vector of measurement data $\V{g}$ is of size $(16 \times 1)$. In order to vary the structure's bending, a square metal profile with feed size $h=20 \unit{mm}$ is placed between the structure and the supporting mounting platforms at the $\xi_i$ positions. The results for these tests are shown in Fig.~\ref{fig:inclinometerCompare1} to \ref{fig:inclinometerError3}. Note, that the reference data has been acquired with calipers and hardly represents the true value; however, it is a good basis for comparisons. 
The constrained linear differential operator $\M{M}$ is of size $(16 \times 16)$. $\M{M}$ and the homogeneous solution $y_h$ are computed in a preparatory step. The online computation is carried out by a BeagleBone Black. The hardware features a RISC processor based on the ARMv7 Cortex A8 platform with $1 \unit{GHz}$ ($2000$ MIPS) and $512 \unit{MB}$ memory. The measurement data $\V{g}$ is acquired via the built-in RS232 serial interface, the results $\V{y}$ are transmitted to a centralized database. The WAGO PFC-200 would be an industrial-ready alternative hardware solution with similar architecture for this application. It features a variety of bus interfaces such as Modbus, Profibus and CAN bus.
}
\section{Conclusion and Outlook}
It can be concluded, from the numerical and experimental tests, that the newly proposed algebraic method outperforms previous solutions, both in accuracy and speed, for the class of problems being considered. 
The separation of the computation into an initial preparatory and a cyclic run-time portion yields a highly efficient numeric solution. 
The computation complexity of the explicit solution is only a function of the number of nodes (sensors) used. 
The automatic generation of C code, and the verification of its correct functionality on multiple embedded architectures has been demonstrated.
The generation of C code also facilitates the use of the method in conjunction with commercial programmable logic controllers (PLCs), for the control of industrial plants and machinery.
Here, the method was applied to a linear array of sensors.
Presently, the tools are being extended to two-dimensional arrays and the resulting two-dimensional fields of data.
\bibliographystyle{ACM-Reference-Format-Journals}
\bibliography{acmtecsgugg2014}
\end{document}

%% file: myDefinitions.tex
\newcommand{\me}{\mathrm{e}}

\newcommand{\defas}{\triangleq}

\newcommand{\M}[1]{\ensuremath{\mathsf{#1}}}
\newcommand{\MT}[1]{\ensuremath{\mathsf{#1}}^\textrm{T}}

\newcommand{\V}[1]{\ensuremath{\boldsymbol{#1}}}

\newcommand{\VO}{\textrm{\textit{\textbf{1}}}} 

\newcommand{\VZ}{\textrm{\textit{\textbf{0}}}} 

\newcommand{\eqtext}[1]{\text{\hspace{3mm} #1 \hspace{3mm}} }
\newcommand{\unit}[1]{\,\ensuremath{[\textnormal{\textrm{#1}}]}}
\newcommand{\rank}[1]{\,\mathrm{rank}\left\{\,#1\,\right\}}
\newcommand{\range}[1]{\,\mathrm{range}\left\{\,#1\,\right\}}

\newcommand{\nullS}[1]{\mathop{\mathrm{null}}\left\{#1\right\}}

\newcommand{\spanS}[1]{\mathop{\mathrm{span}}\left\{#1\right\}}

\renewcommand{\P}[1]{\ensuremath{\mathsf{#1}}}